\documentclass[prd,nofootinbib,preprint,showpacs,showkeys,superscriptaddress]{revtex4}

\usepackage[english]{babel}
\usepackage{amsmath,amssymb}
\usepackage[dvips]{graphicx}
\usepackage{ifthen,array}

\allowdisplaybreaks

\newcommand{\eg} {{\it e.g.}}

\newcommand{\CL}   {C.L.}
\newcommand{\dof}  {d.o.f.}
\newcommand{\eVq}  {\text{eV}^2}
\newcommand{\Sol}  {\textsc{sol}}
\newcommand{\SlKm} {\textsc{sol+kam}}
\newcommand{\Atm}  {\textsc{atm}}

\newcommand{\Dms}  {\Delta m^2_\Sol}
\newcommand{\Dma}  {\Delta m^2_\Atm}
\newcommand{\Dcq}  {\Delta\chi^2}
\newcommand{\EtAl}  {{\it et al.\/}}

\newcommand{\JSQ}   {Just-So$^2$}
\newcommand{\snocc} {SNO$_\mathrm{CC}^\mathrm{rate}$}
\newcommand{\snotot}{SNO$_\mathrm{CC,NC}^\mathrm{SP,DN}$}

\newcommand{\flux}[2][]{\ensuremath{\ifthenelse{\equal{#1}{}}{}{^{#1}\!}\mathit{#2}}}

\newcommand{\AddrAHEP}{%
  Instituto de F\'{\i}sica Corpuscular --
  C.S.I.C./Universitat de Val{\`e}ncia \\
  Edificio Institutos de Paterna, Apt 22085,
  E--46071 Valencia, Spain}

\newcommand{\AddrWIEN}{%
  Institut f\"ur Theoretische Physik, Universit\"at Wien\\
  Boltzmanngasse 5, A--1090 Wien, Austria}

\begin{document}

\preprint{hep-ph/0207227}
\preprint{IFIC/02-26}
\preprint{UWThPh-2002-22}

\title{Constraining neutrino oscillation parameters with current solar
  and atmospheric data}

\author{M.~Maltoni} \email{maltoni@ific.uv.es}
\affiliation{\AddrAHEP}

\author{T.~Schwetz} \email{schwetz@thp.univie.ac.at}
\affiliation{\AddrWIEN}

\author{M.~A.~T{\'o}rtola} \email{mariam@ific.uv.es}
\author{J.~W.~F.~Valle} \email{valle@ific.uv.es}
\affiliation{\AddrAHEP}

\begin{abstract}
    We analyze the impact of recent solar and atmospheric data in the
    determination of the neutrino oscillation parameters, taking into
    account that both the solar $\nu_e$ and the atmospheric $\nu_\mu$
    may convert to a mixture of active and sterile neutrinos. We use
    the most recent global solar neutrino data, including the 1496-day
    Super-K neutrino data sample, and we investigate in detail the
    impact of the recent SNO neutral current, spectral and day/night
    data by performing also an analysis using only the charged current
    rate from SNO. We confirm the clear preference of the pure active
    LMA solution of the solar neutrino problem and obtain that the
    LOW, VAC, SMA and \JSQ\ solutions are disfavored with a $\Dcq =
    9$, $9$, $23$, $31$, respectively.  Furthermore, we find that the
    global solar data constrains the admixture of a sterile neutrino
    to be less than 44\% at 99\% \CL. A pure sterile solution is ruled
    out with respect to the active one at 99.997\% \CL. By performing
    an improved fit of the atmospheric data, we also update the
    corresponding regions of oscillation parameters.  We find that the
    recent atmospheric Super-K (1489-day) and MACRO data have a strong
    impact on constraining a sterile component in atmospheric
    oscillations: if the $\nu_\mu$ is restricted to the atmospheric
    mass states only a sterile admixture of 16\% is allowed at 99\%
    \CL, while a bound of 35\% is obtained in the unconstrained case.
    Pure sterile oscillations are disfavored with a $\Dcq = 34.6$
    compared to the pure active case.
    
    In the appendix we discuss the implications of the first 145.1
    days of KamLAND data on the determination of the solar neutrino
    parameters.  The inclusion of KamLAND enhances the rejection of
    non-LMA-MSW solutions by 13 units in $\Dcq$.  The bound on the
    sterile neutrino fraction is practically unaffected in the
    boron-fixed case, while it improves from 61\% to 51\% in the
    boron-free case.
\end{abstract}



\maketitle

\section{Introduction}

Apart from confirming, yet again, the long-standing solar neutrino
problem~\cite{sksol,chlorine,sage,gallex_gno,sno}, the recent results
from the Sudbury Neutrino Observatory (SNO) on neutral current (NC)
events~\cite{Ahmad:2002jz,Ahmad:2002ka} have given strong evidence
that solar neutrinos convert mainly to an active neutrino flavor. In
addition, valuable spectral and day/night information has been
provided~\cite{Ahmad:2002jz,Ahmad:2002ka}.
This adds to the already robust evidence that an extension of the
Standard Model of particle physics is necessary in the lepton sector.
Although certainly not yet unique, at least for the case of solar
neutrinos, which can be accounted well by spin-flavor
precession~\cite{Miranda:2000bi,Barranco:2002te} or non-standard
neutrino matter interactions~\cite{Guzzo:2001mi}, the most popular
joint explanation of solar and atmospheric experiments is provided by
the neutrino oscillations hypothesis, with neutrino mass-squared
differences of the order of $\Dms\lesssim 10^{-4}~\eVq$ and $\Dma\sim
3 \times 10^{-3}~\eVq$, respectively.

In the wake of the recent SNO NC results we have re-analyzed the
global status of current neutrino oscillation data including these and
the remaining solar
data~\cite{sksol,chlorine,sage,gallex_gno,sno,Ahmad:2002jz,Ahmad:2002ka}
as well as the current atmospheric~\cite{atm-exp,macroOld} samples,
including the 1489 days Super-Kamiokande data~\cite{skatm} and the
most recent MACRO data~\cite{macroNew}.
Motivated by the stringent limits from reactor
experiments~\cite{CHOOZ} we adopt an effective two-neutrino approach
in which solar and atmospheric analyses decouple. However our
effective two-neutrino approach is generalized in the sense that it
takes into account that a light sterile
neutrino~\cite{ptv92,pv93,cm93,Ioannisian:2001mu,Hirsch:2000xe,giuntiwebp},
advocated to account for the LSND anomaly~\cite{LSND}, may take part
in both solar and atmospheric conversions.  The natural setting for
such a light sterile neutrino is provided by four-neutrino models. In
this paper we will determine the constraints on oscillation parameters
in this generalized scenario following from solar and atmospheric data
separately. Such separate analyses are necessary ingredients towards a
combined analysis of all current oscillation data, including solar,
atmospheric, negative short-baseline data and the LSND
experiment~\cite{Maltoni:2001bc,NewFour}. As shown in
Ref.~\cite{Maltoni:2001bc} such separate analyses can be performed
independently of the details of the four-neutrino mass scheme.

Since the release of the latest SNO data in April 2002 a number of
global solar neutrino analyses in terms of active oscillations
appeared~\cite{Barranco:2002te,Bahcall,Band,Barger,Aliani,SMI,update,Fogli:2002pt}.
Moreover, it has been shown by model-independent comparisons of the
SNO CC rate with the SNO NC and Super-K rates that transitions of
solar neutrinos into sterile neutrinos are strongly constrained by the
recent data (see, \eg, Refs.~\cite{Ahmad:2002jz,Band,Barger,Aliani}).
However, so-far no dedicated global analyses exist, where a
participation of a sterile neutrino in the oscillations is fully taken
into account\footnote{In Ref.~\cite{Bahcall:2002zh} admixtures of a
sterile neutrino to solar oscillations are considered. However, the
authors of Ref.~\cite{Bahcall:2002zh} are mainly interested in the
determination of the solar neutrino fluxes and hence, their results
are complementary to those obtained here. Some considerations of
sterile solar neutrino oscillations can also be found in
Ref.~\cite{update}.}. Here we present a complete solar neutrino
analysis including sterile neutrinos, determining the allowed ranges
for the oscillation parameters $\theta_\Sol$ and $\Dms$, as well as
for the parameter $0\le \eta_s \le 1$ describing the active-sterile
admixture. Furthermore, we investigate in detail the impact of the SNO
neutral current, spectral and day/night data and compare with an
analysis where we use only the charged current rate from SNO.

Concerning the atmospheric data, we perform an update of previous
analyses~\cite{Maltoni:2001bc,concha4nu}, adopting again the most
general parameterization of atmospheric neutrino oscillations in the
presence of sterile neutrino mixing, characterized by four parameters.
We find that the recent 1489-day Super-Kamiokande data combined with
the latest MACRO data lead to considerably stronger rejection against
a sterile neutrino contribution to the oscillations than the previous
1289-day data sample.

The plan of the paper is as follows. In Sec.~\ref{sec:sol-oscill} we
set the general parametrization for solar oscillations in the presence
of active-sterile mixing. In Sec.~\ref{sec:sol-analysis} we
briefly describe the solar neutrino data and their analysis.
In Sec.~\ref{sec:sol-results} we present the results of
our analysis, aimed at studying the impact of recent solar data in the
determination of the solar neutrino oscillation parameters, assuming,
as mentioned, that the $\nu_e$ may convert to a mixture of active and
sterile neutrinos.
We give the regions of oscillation parameters for different allowed
$\eta_s$ values, display the global behavior of $\Dcq_\Sol(\Dms)$ and
$\Dcq_\Sol(\theta_\Sol)$, calculated with respect to the favored
active LMA solution, and evaluate the impact of the SNO NC, spectral
and day/night data.
Present solar data exhibit a higher degree of rejection against
non-LMA and/or non-active oscillation solutions, which we quantify,
giving also the absolute goodness of fit (GOF) of various oscillation
solutions.
Our solar neutrino results are briefly compared with those obtained in
other recent analyses in Sec.~\ref{sec:sol-compare}.
In Sec.~\ref{sec:atm-oscill} we set our notations for atmospheric
oscillations in the presence of active-sterile admixture. In
Sec.~\ref{sec:atm-analysis} we briefly describe the
atmospheric neutrino data and their analysis.
In Sec.~\ref{sec:atm-results} we describe our results for
atmospheric oscillation parameters in an improved global fit of
current atmospheric neutrino data. We quantify the impact both of our
improved analysis as well as that of the recent data in rejecting
against the sterile oscillation hypothesis.
We update the corresponding regions of oscillation parameters and
display the global behavior of $\Dcq_\Atm(\Dma)$ and
$\Dcq_\Atm(\theta_\Atm)$. We compare the situation before-and-after
the recent 1489-day atmospheric Super-K data samples and give the
present GOF of the oscillation hypothesis.
In Sec.~\ref{sec:atm-compare}, we briefly compare our atmospheric
neutrino results with those of other analyses.
Finally, in Sec.~\ref{sec:conclusions} we present our conclusions.

\section{Solar neutrinos}
\label{sec:solar}

\subsection{Active-sterile solar neutrino oscillations}
\label{sec:sol-oscill}

In the following we will analyze solar neutrino data in the general
framework of mixed active-sterile neutrino oscillations. In this case
the electron neutrino produced in the sun converts into a combination
of an active non-electron neutrino $\nu_x$ (which again is a
combination of $\nu_\mu$ and $\nu_\tau$) and a sterile neutrino
$\nu_s$:
\begin{equation}
    \nu_e \to \sqrt{1-\eta_s}\, \nu_x + \sqrt{\eta_s}\, \nu_s \,.
\end{equation}
The parameter $\eta_s$ with $0\le \eta_s \le 1$ describes the fraction
of the sterile neutrino participating in the solar oscillations.
Therefore, the oscillation probabilities depend on the three
parameters $\Dms$, $\theta_\Sol$ and $\eta_s$.  The natural framework
of light sterile neutrinos participating in oscillations are
four-neutrino mass schemes, proposed to account for the LSND
result~\cite{LSND} in addition to solar and atmospheric neutrino
oscillations. For previous studies of solar neutrino oscillation in a
four-neutrino framework see
Refs.~\cite{giuntiwebp,concha4nu,giunti4nu} and for an exact
definition of the solar parameters and adopted approximations see
Ref.~\cite{Maltoni:2001bc}.

\subsection{Data and analysis}
\label{sec:sol-analysis}

As experimental data, we use the solar neutrino rates of the chlorine
experiment Homestake~\cite{chlorine} ($2.56 \pm 0.16 \pm 0.16$~SNU),
the most recent result of the gallium experiments
SAGE~\cite{sage}~($70.8 ~^{+5.3}_{-5.2} ~^{+3.7}_{-3.2}$~SNU) and
GALLEX/GNO~\cite{gallex_gno} ($70.8 \pm 4.5 \pm 3.8$~SNU), as well as
the 1496-days Super-Kamiokande data sample~\cite{sksol} in the form of
44 bins (8 energy bins, 6 of which are further divided into 7 zenith
angle bins).
In addition to this, we include the latest results from SNO presented
in Refs.~\cite{Ahmad:2002jz,Ahmad:2002ka}, in the form of 34 data bins
(17 energy bins for each day and night period). Therefore, in our
statistical analysis we use $3+44+34=81$ observables, which we fit in
terms of the three parameters $\Dms$, $\theta_\Sol$ and $\eta_s$, with
a $\chi^2_\Sol$ of the form
\begin{equation}
    \label{eq:1}
    \chi^2_\Sol(\Dms, \theta_\Sol, \eta_s) = \sum_{i,j=1}^{81}
    (R_i^{ex} - R_i^{th}) \cdot
    (\sigma_{ex}^2 + \sigma_{th}^2)_{ij}^{-1} \cdot
    (R_j^{ex} - R_j^{th}) \, .
\end{equation}

In order to fully isolate the impact of the recent neutral current,
spectral and day/night information of the SNO result, we also present
an analysis which does not include such information.  To this aim we
use only the SNO events with energy higher than 6.75~MeV, for which
the NC component is negligible~\cite{sno}.  We sum these events to a
single rate, combining with Cl, Ga rates and full Super-K data, as
described above.  This procedure is analogous to the pre-SNO-NC
situation, except that we take advantage of the enhanced statistics on
the CC rate provided by the new data.  We will refer to this analysis
as \snocc\ analysis and it contains 48 data points.  The comparison
with the analysis including the complete SNO data published this year
(\snotot) allows us to highlight the impact of the SNO NC, spectral
and day/night information.

For the solar neutrino fluxes we use the Standard Solar Model (SSM)
flux~\cite{ssm}, including its standard \flux[8]{B} flux
prediction\footnote{We choose not to include the flux indicated by the
recent $S_{17}$ measurement of Ref.~\cite{Junghans:2001ee}.}.
Motivated by the excellent agreement of the recent SNO NC result with
the predictions of the Standard Solar Model, we prefer to adopt a
boron-fixed analysis. However, for case of the LMA solution we
explicitly illustrate the effect of this assumption by performing also
a boron-free analysis, where we treat the solar \flux[8]{B} flux as
free parameter in the fit.  For simplicity we neglect the \flux{hep}
and \flux{F} neutrino fluxes, whose contribution to the present solar
neutrino experiments is marginal, while for the \flux{pp}, \flux{Be},
\flux{B}, \flux{pep}, \flux{N} and \flux{O} fluxes we use the SSM
value given in Ref.~\cite{ssm}, taking properly into account their
theoretical uncertainties and cross-correlations in the calculation of
the $\chi^2$ function.

For the neutrino cross sections of Chlorine, SAGE, GALLEX/GNO and
Super-K we assume the same as used in previous
papers~\cite{Gonzalez-Garcia:1999aj,GMPV,bahcallNew}, while for the CC
and NC neutrino deuteron differential cross sections relevant for SNO
we use the tables given in~\cite{kubodera}. The contribution of the
cross-section uncertainties to the covariance matrix for the Chlorine
and Gallium experiments is calculated as suggested in
Ref.~\cite{Bahcall:2002zh}. For a given experiment (Chlorine or
Gallium) we use full correlation of the error on the cross section for
low-energy neutrino fluxes (\flux{pp}, \flux{pep}, \flux{Be}, \flux{N} and
\flux{O}), but no correlation of the cross section error between the
low-energy fluxes and the higher-energy \flux[8]{B} flux.

The neutrino survival probability $P_{ee}$ is extracted from the
neutrino evolution operator $\mathbf{U}$, which we factorize as a
product of three factors $\mathbf{U}_\mathrm{sun}$,
$\mathbf{U}_\mathrm{vac}$ and $\mathbf{U}_\mathrm{earth}$
corresponding to propagation in the Sun, vacuum, and Earth,
respectively. The first and last factors include matter effects with
the corresponding density profiles given in Refs.~\cite{ssm}
and~\cite{PREM}. As a simplifying approximation, we assume that
$\mathbf{U}_\mathrm{sun}$ depends only on the neutrino production
point $\vec{x}_0$, $\mathbf{U}_\mathrm{vac}$ only on the Sun-Earth
distance $L$ and $\mathbf{U}_\mathrm{earth}$ depends only on the
zenith-angle $\zeta$ of the incoming neutrinos. Therefore in our
calculations we neglect the small correlation between seasonal effects
and day-night asymmetry~\cite{deHolanda:1999ty}. For each value of the
neutrino oscillation parameters $\Dms/E$, $\theta_\Sol$ and $\eta_s$
we calculate the neutrino survival probability $P_{ee}$ by averaging
over $\vec{x}_0$, $L$ and $\zeta$, properly accounting for all the
interference terms between $\mathbf{U}_\mathrm{sun}$,
$\mathbf{U}_\mathrm{vac}$ and $\mathbf{U}_\mathrm{earth}$.

Special care is taken in including all the theoretical and
experimental errors and their cross-correlations in the calculation
of the covariance matrix, for which we follow the description of
Ref.~\cite{Fogli:2002pt} (covariance approach). In particular, the
errors associated to the Boron-flux shape, the energy-scale and the
energy-resolution uncertainties of the Super-Kamiokande and SNO
experiments are recalculated for each point in parameter space.

\subsection{Results and discussion}
\label{sec:sol-results}

In order to determine the expected event numbers for the various solar
neutrino experiments we calculate the $\nu_e$ survival probability for
each point in parameter space of ($\tan^2\theta_\Sol,\Dms,\eta_s$) and
convolute it with the Standard Solar Model neutrino fluxes~\cite{ssm}
and the relevant neutrino cross sections. We have compared such
expected event numbers with the data described above, taking into
account the detector characteristics and appropriate response
functions.  Using the above-mentioned $\chi^2_\Sol$ we have performed
a global fit of solar neutrino data, whose results we now summarize.

\begin{figure}[t] \centering
    \includegraphics[width=0.95\linewidth]{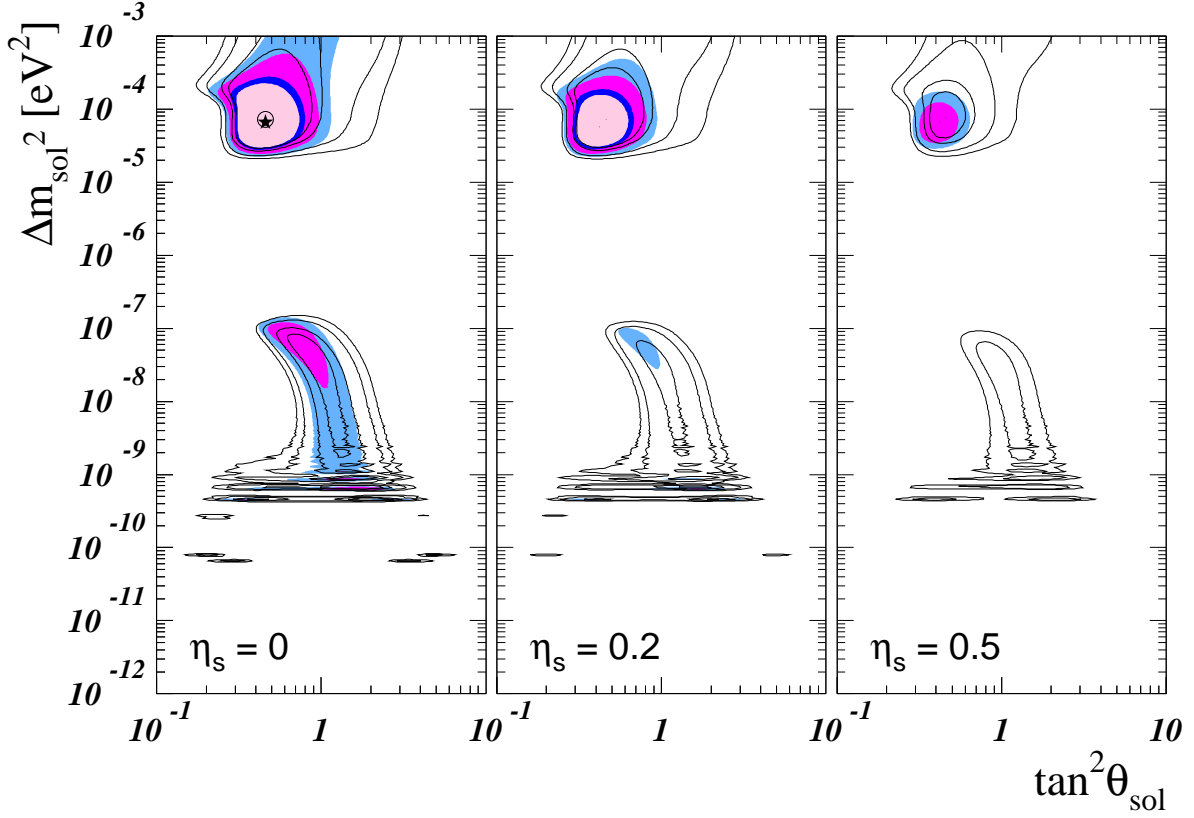}
    \caption{\label{fig:sol-region}%
      Allowed regions of $\tan^2\theta_\Sol$ and $\Dms$ for $\eta_s =
      0$ (active oscillations), $\eta_s = 0.2$ and $\eta_s = 0.5$. The
      lines indicate the regions determined by the \snocc\ analysis
      (see definition in text), the shaded regions correspond to
      \snotot\ (see text). The confidence levels are 90\%, 95\%, 99\%
      and 3$\sigma$ for 3 \dof.}
\end{figure}

Our global best-fit point occurs for the values
\begin{equation} \label{eq:bfp-sol}
    \tan^2\theta_\Sol = 0.46, \qquad \Dms = 6.6\times 10^{-5}~\eVq
\end{equation}
and correspond to $\eta_s=0$. We obtain a $\chi^2_\mathrm{min}=65.8$
for $81-3$ \dof, leading to the excellent goodness of fit of 84\%.
In Fig.~\ref{fig:sol-region} we display the regions of solar neutrino
oscillation parameters for 3 \dof\ with respect to this global
minimum, for the standard case of active oscillations, $\eta_s = 0$,
as well as for $\eta_s = 0.2$ and $\eta_s = 0.5$.
The first thing to notice is the impact of the SNO NC, spectral, and
day/night data in improving the determination of the oscillation
parameters: the shaded regions after their inclusion are much smaller
than the hollow regions delimited by the corresponding \snocc\
confidence contours. Especially important is the full \snotot\
information for excluding {\it maximal} solar mixing in the LMA region
and in closing the LMA region from above in $\Dms$.  Values of $\Dms >
10^{-3}~\eVq$ appear only at $3\sigma$. Previously solar data on its
own could not close the LMA region, only the inclusion of data from
reactor experiments~\cite{CHOOZ} ruled out the upper part of the LMA
region~\cite{GMPV}. We obtain the following $3\sigma$ ranges (1 \dof):
\begin{equation} \label{eq:sol_ranges}
    \text{LMA:}\qquad
    0.26 \le \tan^2\theta_\Sol \le 0.85, \qquad
    2.6\times 10^{-5}~\eVq \le \Dms \le
    3.3\times 10^{-4}~\eVq.
\end{equation}
It is interesting to note that these $3\sigma$ intervals are
essentially unchanged if we minimize with respect to $\eta_s$ or if we
apply the constraint $\eta_s=0$ (pure active oscillations).
In order to compare our allowed regions given in
Fig.~\ref{fig:sol-region} with those of other groups, one has to take
into account that we calculate the \CL\ regions for the 3 \dof\
$\tan^2\theta_\Sol$, $\Dms$ and $\eta_s$. Therefore at a given \CL\
our regions are larger than the usual regions for 2 \dof, because we
also constrain the parameter $\eta_s$.

\begin{figure}[t] \centering
    \includegraphics[width=0.95\linewidth]{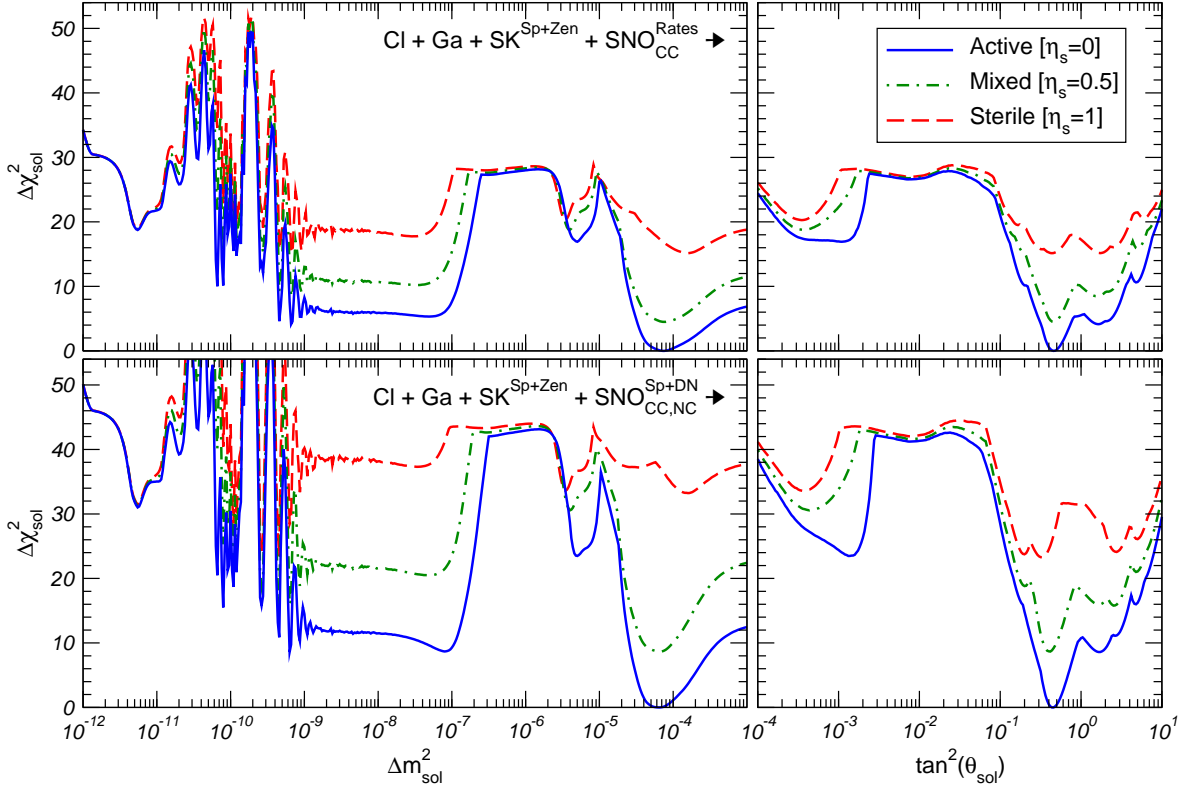}
    \caption{ \label{fig:sol-chisq}%
      $\Dcq_\Sol$ as a function of $\Dms$ and
      $\tan^2\theta_\Sol$, for pure active ($\eta_s = 0$), pure
      sterile ($\eta_s = 1$) and mixed neutrino oscillations ($\eta_s
      = 0.5$). Upper and lower panels correspond to the \snocc\ and
      \snotot\ samples defined in text.}
\end{figure}

\begin{figure}[t] \centering
    \includegraphics[width=0.95\linewidth]{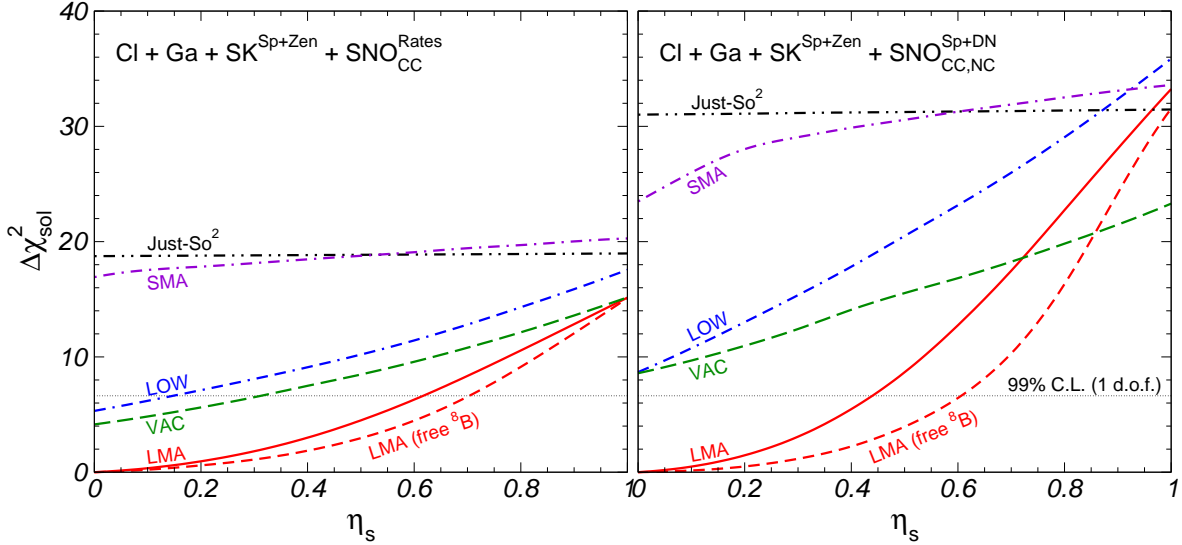}
    \caption{ \label{fig:sol-etas}%
      $\Dcq_\Sol$ displayed as a function of $\eta_s$ with
      respect to favored active LMA solution, for the \snocc\ (left
      panel) and the \snotot\ (right panel) analysis, as defined in
      text.}
\end{figure}

Next we notice the enhanced discrimination against non-LMA solutions
implied by the new data, apparent in Figs.~\ref{fig:sol-region},
\ref{fig:sol-chisq} and \ref{fig:sol-etas}. This shows that the first
hints~\cite{Gonzalez-Garcia:1999aj,Bahcall:1999ed} in favor of a
globally preferred LMA oscillation solution which followed mainly from
the flatness of the Super-K spectra, have now become a robust result,
thanks to the additional data, to which SNO has contributed
significantly\footnote{See also Ref.~\cite{strumiaLMA}.}.  One sees
that, in contrast with the \snocc\ situation, non-LMA solutions do not
appear at 95\% \CL.  However, the LOW and VAC solutions still appear
at 99\% \CL\ for 3 \dof.

In order to concisely illustrate the above results we display in
Fig.~\ref{fig:sol-chisq} the profiles of $\Dcq_\Sol$ as a function of
$\Dms$ (left) as well as $\tan^2\theta_\Sol$ (right), by minimizing
with respect to the undisplayed oscillation parameters, for the fixed
values of $\eta_s=0$, $0.5$, $1$.
By comparing top and bottom panels in Fig.~\ref{fig:sol-chisq} one can
clearly see the impact of the full \snotot\ sample in leading to the
relative worsening of all non-LMA solutions with respect to the
preferred active LMA solution.


\begin{table}[t] \centering\small
    \catcode`?=\active \def?{\hphantom{0}}
    \newcommand{\E}[2]{${#1}\times 10^{#2}$}
    \begin{tabular}{|>{\rule[-2mm]{0pt}{6mm}}l|cccc|cccc|}
        \hline
        & \multicolumn{4}{c|}{\snocc} & \multicolumn{4}{c|}{\snotot}
        \\
        \hline
        Region
        & $\tan^2\theta_\Sol$ & $\Delta m_\Sol^2$ & $\chi_\Sol^2$ & GOF
        & $\tan^2\theta_\Sol$ & $\Delta m_\Sol^2$ & $\chi_\Sol^2$ & GOF
        \\
        \hline
        \multicolumn{9}{|>{\rule[-2mm]{0pt}{6mm}}c|}{Pure active ($\eta_s = 0$)} \\
        \hline
        LMA  & 0.46        & \E{7.2}{-5?} & 40.9 & 69\% & 0.46        & \E{6.6}{-5?} & ?65.8 & 86\% \\
        LOW  & 0.83        & \E{4.8}{-8?} & 46.2 & 46\% & 0.66        & \E{7.9}{-8?} & ?74.4 & 62\% \\
        VAC  & 1.7?        & \E{6.6}{-10} & 45.0 & 51\% & 1.7?        & \E{6.3}{-10} & ?74.4 & 63\% \\
        SMA  & \E{1.1}{-3} & \E{5.0}{-6?} & 57.8 & 11\% & \E{1.4}{-3} & \E{5.0}{-6?} & ?89.3 & 20\% \\
        \JSQ & 1.0?        & \E{5.5}{-12} & 59.6 & ?9\% & 1.0?        & \E{5.5}{-12} & ?96.8 & ?8\% \\
        \hline
        \multicolumn{9}{|>{\rule[-2mm]{0pt}{6mm}}c|}{Mixed ($\eta_s = 0.5$)} \\
        \hline
        LMA  & 0.46        & \E{7.6}{-5?} & 45.4 & 50\% & 0.42        & \E{6.6}{-5?} & ?74.4 & 62\% \\
        LOW  & 0.91        & \E{3.5}{-8?} & 51.1 & 28\% & 0.83        & \E{4.8}{-8?} & ?86.3 & 27\% \\
        VAC  & 1.6?        & \E{6.9}{-10} & 49.4 & 34\% & 0.35        & \E{4.6}{-10} & ?81.3 & 41\% \\
        SMA  & \E{3.6}{-4} & \E{4.0}{-6?} & 59.7 & ?8\% & \E{4.4}{-4} & \E{4.0}{-6?} & ?96.3 & ?9\% \\
        \JSQ & 1.0?        & \E{5.5}{-12} & 59.8 & ?8\% & 1.0?        & \E{5.5}{-12} & ?97.0 & ?8\% \\
        \hline
        \multicolumn{9}{|>{\rule[-2mm]{0pt}{6mm}}c|}{Pure sterile ($\eta_s = 1$)} \\
        \hline
        LMA  & 0.44        & \E{1.6}{-4?} & 56.0 & 15\% & 0.38        & \E{1.6}{-4?} & ?99.0 & ?6\% \\
        LOW  & 1.6?        & \E{1.4}{-9?} & 58.5 & 10\% & 1.6?        & \E{1.1}{-9?} & 101.6 & ?4\% \\
        VAC  & 1.7?        & \E{6.9}{-10} & 56.1 & 15\% & 0.33        & \E{4.6}{-10} & ?89.1 & 21\% \\
        SMA  & \E{3.5}{-4} & \E{3.5}{-6?} & 61.2 & ?7\% & \E{3.6}{-4} & \E{3.5}{-6?} & ?99.4 & ?6\% \\
        \JSQ & 1.1?        & \E{5.5}{-12} & 59.9 & ?8\% & 1.0?        & \E{5.5}{-12} & ?97.2 & ?8\% \\
        \hline
    \end{tabular}
    \caption{ \label{tab:sol-chisq}%
      Best-fit values of $\Dms$ and $\theta_\Sol$ with the
      corresponding $\chi^2_\Sol$ and GOF for pure active, pure
      sterile, and mixed neutrino oscillations.  Results are given for
      the \snocc\ (left column) and for the full \snotot\ analysis
      (right column). The relevant number of \dof\ is $48-2$ ($81-2$)
      for the \snocc\ (\snotot) analysis. }
\end{table}

The corresponding best-fit values for the various solutions of $\Dms$
and $\theta_\Sol$ and the values of $\chi^2_\Sol$ evaluated at the
best-fit points are compiled in Tab.~\ref{tab:sol-chisq}. This table
gives results for the three cases considered above: pure active, pure
sterile and mixed neutrino oscillations, both for the \snocc\ and the
full \snotot\ analysis.  To calculate the goodness of fit of the
various solutions we evaluate in this table the $\chi^2$ for $48-2$
$(81-2)$ \dof\ for the \snocc\ (\snotot) analysis defined previously.
Note that we fix $\eta_s$ at the three values $0$, $0.5$ and $1$. In
the pure active case we find for LOW, VAC, SMA and \JSQ\ the following
differences in $\chi^2$ relative to the global best-fit point in LMA
\begin{equation} \label{eq:dcq}
    \Dcq_{\text{LOW}}  =  8.7, \qquad
    \Dcq_{\text{VAC}}  =  8.6, \qquad
    \Dcq_{\text{SMA}}  = 23.5, \qquad
    \Dcq_{\text{\JSQ}} = 31.0.
\end{equation}
Note that especially SMA and \JSQ\ are highly disfavored with respect
to LMA.


In addition to the scrutiny of the different neutrino oscillation
solutions in the solar neutrino oscillation parameters $\Dms$ and
$\theta_\Sol$, the present solar data can test the sterile neutrino
oscillation hypothesis, characterized by the parameter $\eta_s$
introduced above. The results can be presented in several equivalent
ways.  For example, rejection of sterile solar neutrino oscillations
is already hinted by comparing the middle and right panels of
Fig.~\ref{fig:sol-region} with the left one, corresponding to the
pure active oscillation case: clearly the solutions deteriorate as
$\eta_s$ increases. Furthermore, the lines for $\eta_s=0.5$ and
$\eta_s=1$ shown in Fig.~\ref{fig:sol-chisq} clearly show that sterile
solutions are strongly disfavored with respect to pure active
solutions.

In order to summarize the above results we display in
Fig.~\ref{fig:sol-etas} the profile of $\Dcq_\Sol$ as a function of $0
\leq \eta_s \leq 1$, irrespective of the detailed values of the solar
neutrino oscillation parameters $\Dms$ and $\theta_\Sol$. This figure
clearly illustrates the degree with which the solar neutrino data
sample rejects the presence of a sterile component for each one of the
possible solar neutrino oscillation solutions.
The figure shows how the preferred LMA status survives in the presence
of a small sterile component characterized by $\eta_s$ (also seen in
Figs.~\ref{fig:sol-region} and \ref{fig:sol-chisq}). Further, one sees
that the value $\eta_s=0$ is always preferred, so that increasing
$\eta_s$ leads to a deterioration of all oscillation solutions.
Notice that there is a crossing between the LMA and VAC solutions, as
a result of which the best pure sterile description lies in the vacuum
regime. However, in the global analysis pure sterile oscillations with
$\eta_s=1$ are highly disfavored. We find a $\chi^2$-difference
between pure active and sterile of $\Dcq_\mathrm{s-a} = 33.2$ if we
restrict to the LMA solution, or $\Dcq_\mathrm{s-a} = 23.3$ if we
allow also for VAC. For 3 \dof\ the $\Dcq_\mathrm{s-a} = 23.3$ implies
that pure sterile oscillations are ruled out at 99.997\% \CL\ compared
to the active case.

For the LMA solution we have also performed an analysis without fixing
the boron flux to its SSM prediction. In this case we treat the
\flux[8]{B} flux as a free parameter in the fit, and remove the error
on this flux from the covariance matrix.  From Fig.~\ref{fig:sol-etas}
one can see that the constraint on $\eta_s$ is weaker in the
boron-free case than in the boron-fixed one, since a {\it small}
sterile component can now be partially compensated by increasing the
total \flux[8]{B} flux coming from the Sun. From the figure we obtain
the bounds
\begin{equation} \label{eq:etasSol}
    \text{solar data:}
    \qquad \eta_s \leq 0.44 \text{~(boron-fixed)},
    \qquad \eta_s \leq 0.61 \text{~(boron-free)}
\end{equation}
at 99\% \CL\ for 1 \dof. In summary, we have found that, as long as
the admixture of sterile neutrinos is acceptably small, the LMA is
always the best of the oscillation solutions, establishing its
robustness also in our generalized oscillation scheme.

\begin{figure}[t] \centering
    \includegraphics[width=0.9\textwidth]{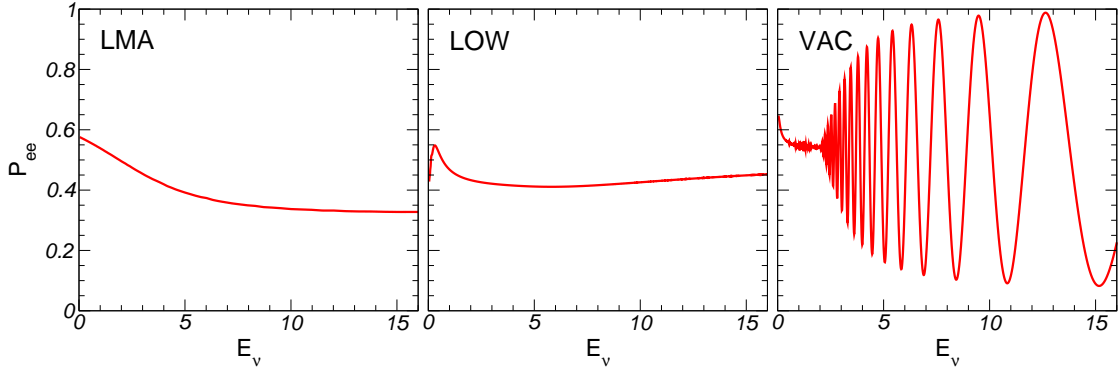}
    \caption{ \label{fig:sol-probab}%
      Best-fit active solar neutrino survival probabilities.}
\end{figure}

To round off our discussion of the solar neutrino fit update we
present in Fig.~\ref{fig:sol-probab} the $\nu_e$ survival probability
versus energy $E$ for the various solutions LMA, LOW and VAC,
calculated as described above at the local $\chi^2$-minima given in
Tab.~\ref{tab:sol-chisq}.
Similar plots can be made for the case of sterile oscillations.


\subsection{Comparison with other groups}
\label{sec:sol-compare}

\begin{table}[!t] \centering\small
    \catcode`?=\active \def?{\hphantom{0}}
    \begin{tabular}{|>{\rule[-2mm]{0pt}{6mm}}l|c|c|c|c|c|c|c|c|c|c|c|}
        \hline
        & \rotatebox{90}{SNO Collaboration~\cite{Ahmad:2002ka}, nucl-ex/0204009~v2, Table 4}
        & \rotatebox{90}{Super-K Collaboration~\cite{sksol}, hep-ex/0205075~v1, Table~3} 
        & \rotatebox{90}{Barger \EtAl~\cite{Barger}, hep-ph/0204253~v3, Table~2} 
        & \rotatebox{90}{Bandyopadhyay \EtAl~\cite{Band}, hep-ph/0204286~v4, Table~2 } 
        & \rotatebox{90}{Bahcall \EtAl~\cite{Bahcall}, hep-ph/0204314~v4, Table~4}
        & \rotatebox{90}{Creminelli \EtAl~\cite{update}, hep-ph/0102234~v3, Chapter 7} 
        & \rotatebox{90}{Aliani \EtAl~\cite{Aliani}, hep-ph/0205053~v3, Table~2} 
        & \rotatebox{90}{De~Holanda, Smirnov~\cite{SMI}, hep-ph/0205241~v3, Table~1} 
        & \rotatebox{90}{Fogli \EtAl~\cite{Fogli:2002pt}, hep-ph/0206162~v1, Table~1} 
        & \rotatebox{90}{Barranco \EtAl~\cite{Barranco:2002te}, hep-ph/0207326~v2, Table~1} 
        & \rotatebox{90}{present analysis} \\ 
        \hline
        \dof & $75-3$ & $~\,~46~\,~$ & $75-3$ & $49-4$ & $80-3$ & $49-2$ &
               $41-4$ & $81-3$ & $81-3$ & $81-2$ & $81-2$ \\
        \hline
        best fit & \multicolumn{11}{c|}{LMA solution} \\
        \hline
        $\tan^2\theta_\Sol$
        & 0.34 & 0.38 & 0.39 & 0.41 & 0.45 & 0.45 & 0.40 & 0.41 & 0.42 & 0.47 & 0.46 \\
        $\Delta m_\Sol^2$ [$10^{-5}~\eVq$]
        & 5.0 & 6.9 & 5.6 & 6.1 & 5.8 & 7.9 & 5.4 & 6.1 & 5.8 & 5.6 & 6.6 \\
        $\chi^2_\mathrm{LMA}$
        & 57.0 & 43.5 & 50.7 & 40.6 & 75.4 & 33.0 & 30.8& 65.2 & 73.4 & 68.0 & 65.8 \\
        GOF
        & 90\% & 58\% & 97\% & 66\% & 53\% & 94\% & 80\% & 85\% & 63\% & 81\% & 86\% \\
        \hline
        $\Delta \chi^2_{\text{LOW, active}}$
        & 10.7 & ?9.0 & ?9.2 & 10.0 & ?9.6 & ?8.1 & -- & 12.4 & 10.0 & -- & ?8.7 \\
        $\Delta \chi^2_{\text{VAC, active}}$
        & --   & 10.0 & 25.6 & 15.5 & 10.1 & 14.? & -- & ?9.7 & ?7.8 & -- & ?8.6 \\
        $\Delta \chi^2_{\text{SMA, active}}$
        & --   & 15.4 & 57.3 & 30.4 & 25.6 & 23.? & -- & 34.5 & 23.5 & -- & 23.5 \\
        \hline
        $\Delta \chi^2_{\text{LMA, sterile}}$
        & --   & --   & --   & --   & --   & 29.? & -- & --   & --   & -- & 33.2 \\
        $\Delta \chi^2_{\text{LOW, sterile}}$
        & --   & --   & --   & --   & --   & --   & -- & --   & --   & -- & 35.9 \\
        $\Delta \chi^2_{\text{VAC, sterile}}$
        & --   & --   & --   & --   & 26.0 & --   & -- & --   & --   & -- & 23.3 \\
        $\Delta \chi^2_{\text{SMA, sterile}}$
        & --   & --   & --   & --   & 39.7 & --   & -- & --   & --   & -- & 33.6 \\
        \hline
    \end{tabular}
    \caption{\label{tab:sol-compare}%
      Comparison of solar neutrino analyses among different groups. We
      show the number of analyzed data points minus the fitted
      parameters, the best-fit values of $\tan^2\theta_\Sol$ and $\Dms$
      for active oscillations and the corresponding $\chi^2$-minima and
      GOF. Further we show the $\Delta\chi^2$ with respect to the best
      fit LMA active solution for various other solutions (active, as
      well as sterile).}
\end{table}

Before turning to the atmospheric neutrino fits let us compare our
solar neutrino results with those of other groups.  Since the release
of the latest SNO data in April 2002 several analyses have appeared.
Taking into account the large amount of experimental input data,
variations in the analysis (such as the construction of the $\chi^2$
function or the treatment of theoretical errors) and the complexity of
the codes involved it seems interesting to compare quantitatively the
outcomes of different analyses.  In Tab.~\ref{tab:sol-compare} we have
compiled some illustrative results of the solar neutrino analyses
performed by the SNO and Super-K
collaborations~\cite{sksol,Ahmad:2002ka}, as well as theoretical
ones~\cite{Barranco:2002te,Bahcall,Band,Barger,Aliani,SMI,update,Fogli:2002pt}.

Generally speaking, on statistical grounds, one expects the
differences in the statistical treatment of the data to have little
impact on the global best-fit parameter values, which lie in the LMA
region for all analyses and are in good agreement. These differences
typically become more visible as one compares absolute values of the
$\chi^2$, and/or as one departs from the best-fit region towards more
disfavored solutions. Aware of this, we took special care to details
such as the dependence of the theoretical errors on the oscillation
parameters, which enter in the covariance matrix characterizing the
Super-K and SNO electron recoil spectra. This way we obtain results
which we consider reliable in the full oscillation parameter space.

In the row labeled ``\dof''\ we show the number of analyzed data
points minus the fitted parameters in each analysis\footnote{Here we
  do not treat $\eta_s$ as a free fit parameter, since we consider
  only the limiting cases $\eta_s=0$ and $1$; this is the reason for
  the number $81-2$ in the present analysis.}. One can see from these
numbers that various groups use different experimental input data, in
particular the spectral and zenith angle information of Super-K and/or
SNO is treated in different ways. Despite obvious differences in the
analyses there is relatively good agreement on the best-fit LMA active
oscillation parameters: the obtained best-fit values for
$\tan^2\theta_\Sol$ are in the range $0.34- 0.47$ and for $\Dms$ they
lie in the interval $(5.0 - 7.9)\times 10^{-5}~\eVq$. There is also
good agreement on the allowed ranges of the oscillation parameters
(not shown in the table). For example, the $3\sigma$ intervals given
in Ref.~\cite{Bahcall} ($0.24 \le \tan^2\theta_\Sol \le 0.89$ and
$2.3\times 10^{-5}~\eVq \le \Dms \le 3.7\times 10^{-4}~\eVq$) and in
Ref.~\cite{SMI} ($\tan^2\theta_\Sol \le 0.84$ and $2.3\times
10^{-5}~\eVq \le \Dms \le 3.6\times 10^{-4}~\eVq$) agree very well
with the ranges given in Eq.~\eqref{eq:sol_ranges}.
However, even for the favored LMA solution, there are some differences
in the GOF of the best-fit LMA solution, ranging from
53\%~\cite{Bahcall} to 97\%~\cite{Barger}, due to differences in the
construction of the $\chi^2$ function by different groups.

There is remarkable agreement on the rejection of the LOW solution
with respect to LMA with a $\Delta\chi^2_{\text{LOW, active}} \approx
10$. Our result for the vacuum solution $\Delta\chi^2_{\text{VAC,
    active}} = 8.6$ is in good agreement with the values obtained in
Refs.~\cite{sksol,Bahcall,SMI,Fogli:2002pt}, whereas
Refs.~\cite{Band,Barger,update} obtain higher values.  Our result for
the SMA solution of $\Delta\chi^2_\text{SMA, active} = 23.5$ is in
good agreement with the values obtained in
Refs.~\cite{Bahcall,update,Fogli:2002pt}; while Refs.~\cite{Band,SMI}
and especially Ref.~\cite{Barger} obtain higher values. On the other
hand in Ref.~\cite{sksol} SMA is somewhat less
disfavored\footnote{Tracing back the reason for these and other
  differences in Tab.~\ref{tab:sol-compare} goes beyond the scope of this
  work.}.

There had been so-far no dedicated global analysis of solar neutrino
oscillations including the most recent SNO data for the case
where sterile neutrinos take part in solar oscillations ($\eta_s \neq
0$).  Model-independent considerations of transitions into sterile
neutrinos can be found in Refs.~\cite{Ahmad:2002jz,Band,Barger,Aliani}. Solar
neutrino oscillations in the presence of active-sterile admixtures are
also studied in Ref.~\cite{Bahcall:2002zh}, although in a different
context. In the lower part of Tab.~\ref{tab:sol-compare} we compare the
partial results given in Refs.~\cite{Bahcall} and \cite{update} for
the pure sterile case ($\eta_s=1$) with the corresponding values found
in the present analysis. Although there are noticeable differences of
the shown $\Delta\chi^2$-values, there is agreement on the qualitative
behavior. We have also studied intermediate levels of sterile neutrino
admixture, giving the corresponding regions of oscillation parameters
and the full $\chi^2$ profiles relative to the favored active LMA
solution (not shown in Tab.~\ref{tab:sol-compare}, see
Figs.~\ref{fig:sol-region}, \ref{fig:sol-chisq} and
\ref{fig:sol-etas}).


We now turn to the analysis of the latest atmospheric data. As already
mentioned in the introduction, separate analyses of solar and
atmospheric data samples constitute the necessary ingredients towards
a full combined study of all current oscillation data, including also
the short-baseline data, as shown in~\cite{Maltoni:2001bc,NewFour}.


\section{Atmospheric neutrinos}
\label{sec:atmos}

\subsection{Active-sterile atmospheric neutrino oscillations}
\label{sec:atm-oscill}

In our analysis of atmospheric data we will make use of the hierarchy
$\Dms\ll\Dma$ and neglect the solar mass splitting. Further, in order
to comply with the strong constraints from reactor
experiments~\cite{CHOOZ} we completely decouple the electron neutrino
from atmospheric oscillations\footnote{For a dedicated study of these
  issues see Ref.~\cite{Gonzalez-Garcia:2002mu}.}.
In the following we will consider atmospheric neutrino data in a
generalized oscillation scheme in which a light sterile neutrino takes
part in the oscillations. The setting for such scenarios are
four-neutrino mass schemes~\cite{ptv92,pv93,cm93}. In such schemes,
besides the solar and atmospheric mass-splittings, there is also a
large $\Delta m^2$ motivated by the LSND experiment~\cite{LSND}. In
contrast with the case of solar $\nu_e$ oscillations, the constraints
on the $\nu_\mu$--content in atmospheric oscillations are not so
stringent: in fact such constrains are provided by atmospheric data
themselves~\cite{Maltoni:2001mt}.  As a result to describe atmospheric
neutrino oscillations in this general
framework~\cite{Maltoni:2001bc,concha4nu} we need two more parameters
besides the standard 2-neutrino oscillation parameters $\theta_\Atm$
and $\Dma$. We will use the parameters $d_\mu$ and $d_s$ already
introduced in Ref.~\cite{Maltoni:2001bc}, and defined in such a way
that $(1-d_\mu)$ and $(1-d_s)$ correspond to the fractions of
$\nu_\mu$ and $\nu_s$ participating in oscillations with $\Dma$,
respectively.  Hence, pure active atmospheric oscillations with $\Dma$
are recovered in the limit $d_\mu=0$ and $d_s=1$. In four-neutrino
models there is a mass scheme-dependent relationship between $d_s$ and
the solar parameter $\eta_s$. For details see
Ref.~\cite{Maltoni:2001bc}.

We will also perform an analysis by imposing the constraint $d_\mu=0$.
In such ``restricted'' analysis the $\nu_\mu$ is completely
constrained to the atmospheric mass states. Only in this limit the
parameter $d_s$ has a similar interpretation as $\eta_s$ introduced in
the solar case. For $d_\mu=0$ we obtain that $\nu_\mu$ oscillates into
a linear combination of $\nu_\tau$ and $\nu_s$ with $\Dma$:
\begin{equation}\label{eq:sterileR}
    d_\mu=0: \qquad 
    \nu_\mu \to \sqrt{d_s}\, \nu_\tau + \sqrt{1-d_s}\, \nu_s \,.
\end{equation}

\subsection{Data and analysis}
\label{sec:atm-analysis}

For the atmospheric data analysis we use all the charged-current data
from the Super-Kamiokande~\cite{skatm} and MACRO~\cite{macroNew}
experiments. The Super-Kamiokande data include the $e$-like and
$\mu$-like data samples of sub- and multi-GeV contained events (10
bins in zenith angle), as well as the stopping (5 angular bins) and
through-going (10 angular bins) up-going muon data events. We do not
use the information on $\nu_\tau$ appearance, multi-ring $\mu$ and
neutral-current events since an efficient Monte-Carlo simulation of
these data sample would require a more detailed knowledge of the Super
Kamiokande experiment, and in particular of the way the
neutral-current signal is extracted from the data. Such an information
is presently not available to us. From MACRO we use the through-going
muon sample divided in 10 angular bins~\cite{macroNew}. We did not
include in our fit the results of other atmospheric neutrino
experiments, such as the recent 5.9 kton-yr data from
Soudan-2~\cite{Soudan2}, since at the moment the statistics is
completely dominated by Super-Kamiokande~\cite{GMPV}. Furthermore,
some of the older experiments have no angular sensitivity, and thus
can not be used to discriminate between active and sterile neutrino
conversion, our main goal.

Our statistical analysis of the atmospheric data is similar to that
used in Ref.~\cite{Maltoni:2001bc}, except that we now take advantage
of the new Super-Kamiokande data and of the full ten-bin zenith-angle
distribution for the contained events, rather than the five-bin
distribution employed previously. Therefore, we have now $65$
observables, which we fit in terms of the four relevant parameters
$\Dma$, $\theta_\Atm$, $d_\mu$ and $d_s$:
\begin{equation}
    \label{eq:2}
    \chi^2_\Atm(\Dma, \theta_\Atm, d_\mu, d_s) = \sum_{i,j=1}^{65}
    (N_i^{ex} - N_i^{th}) \cdot
    (\sigma_{ex}^2 + \sigma_{th}^2)_{ij}^{-1} \cdot
    (N_j^{ex} - N_j^{th}) \, .
\end{equation}

Concerning the theoretical Monte-Carlo, we improve the method
presented in Ref.~\cite{GMPV} by properly taking into account the
scattering angle between the incoming neutrino and the scattered
lepton directions. This was already the case for Sub-GeV contained
events, however previously~\cite{Maltoni:2001bc} we made the
simplifying assumption of full neutrino-lepton collinearity in the
calculation of the expected event numbers for the Multi-GeV contained
and up-going-$\mu$ data samples.  While this approximation is still
justified for the stopping and thru-going muon samples, in the
Multi-GeV sample the theoretically predicted value for down-coming
$\nu_\mu$ is systematically higher if full collinearity is assumed.
The reason for this is that the strong suppression observed in these
bins cannot be completely ascribed to the oscillation of the
down-coming neutrinos (which is small due to small travel distance).
Because of the non-negligible neutrino-lepton scattering angle at
these Multi-GeV energies there is a sizable contribution from up-going
neutrinos (with a higher conversion probability due to the longer
travel distance) to the down-coming leptons.
However, this problem is less visible when the angular information of
Multi-GeV events is included in a five angular bins presentation of
the data, as previously assumed~\cite{Gonzalez-Garcia:1998vk}.

\subsection{Results and Discussion}
\label{sec:atm-results}

Folding together the atmospheric neutrino fluxes~\cite{Bartol}, our
calculated neutrino survival probabilities including Earth matter
effects with the profile of Ref.~\cite{PREM}, and the relevant
neutrino cross sections, we determine the expected event numbers for
the various atmospheric neutrino observables, taking into account the
appropriate detector response characteristics.  Comparing with the
data described in Sec.~\ref{sec:atm-analysis}, we have performed a
global fit of the atmospheric neutrino data using the above-discussed
$\chi^2_\Atm$, following the same method used in Ref.~\cite{GMPV}. We
now summarize the main features of this fit.

Our global best-fit point occurs at the parameter values
\begin{equation}
    \sin^2\theta_\Atm = 0.49, \qquad 
    \Dma = 2.1 \times 10^{-3}~\eVq \qquad\text{(best)}
\end{equation}
and $d_s=0.92,\: d_\mu=0.04$. We see that atmospheric data prefers a
small sterile neutrino admixture. However, this effect is not
statistically significant, also the pure active case ($d_s=1,
d_\mu=0$) gives an excellent fit: the difference in $\chi^2$ with
respect to the best-fit point is only $\Dcq_\mathrm{act-best} = 3.3$.
For the pure active best-fit point we obtain
\begin{equation}\label{eq:atmbf}
    \sin^2\theta_\Atm = 0.5, \qquad 
    \Dma = 2.5 \times 10^{-3}~\eVq \qquad\text{(active)}
\end{equation}
with the 3$\sigma$ ranges (1 \dof)
\begin{equation}\label{eq:atm3sigma}
    0.3 \le \sin^2\theta_\Atm \le 0.7, \qquad
    1.2 \times 10^{-3}~\eVq \le \Dma \le 4.8 \times 10^{-3}~\eVq 
    \qquad\text{(active)}.
\end{equation}

\begin{figure}[t] \centering
    \includegraphics[width=0.95\textwidth]{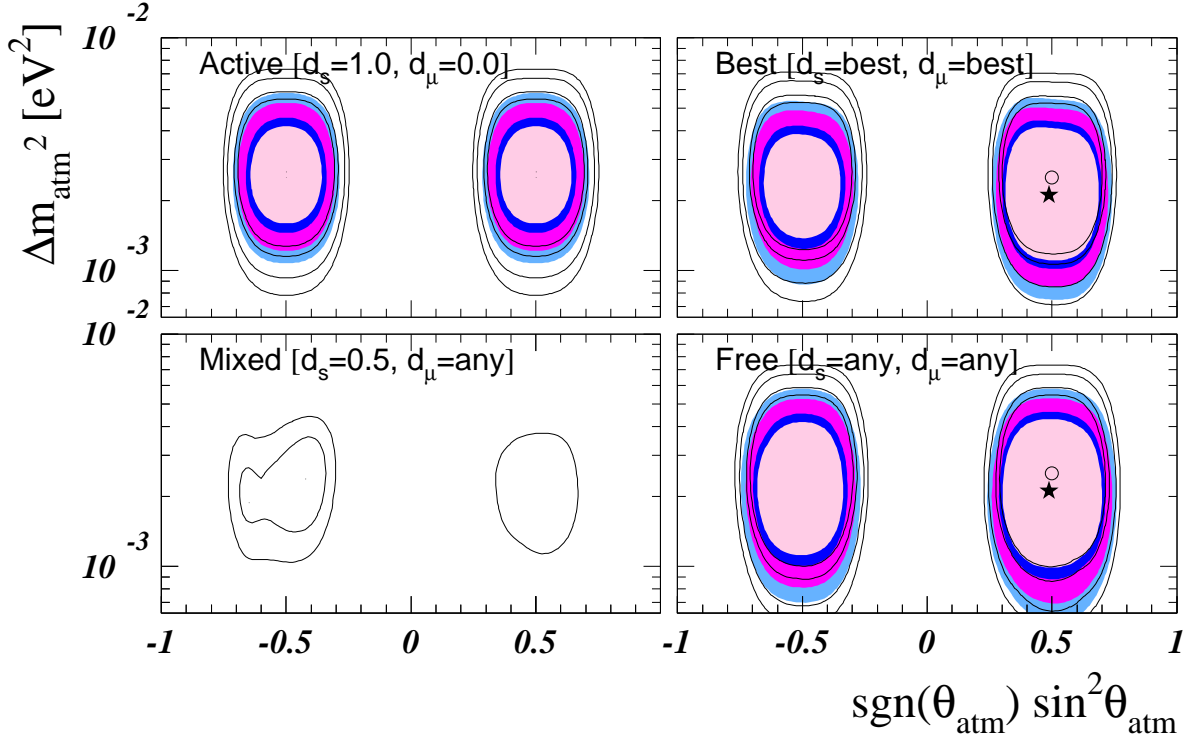}
    \caption{ \label{fig:atm-region}%
      Allowed regions of the parameters $\sin^2\theta_\Atm$ and $\Dma$
      at 90\%, 95\%, 99\% and 3$\sigma$ for 4 \dof\ and different
      assumptions on the parameters $d_s$ and $d_\mu$ (see text). The
      lines (shaded regions) correspond to 1289 (1489) days of Super-K
      data.}
\end{figure}

\begin{figure}[t] \centering
    \includegraphics[width=0.95\linewidth]{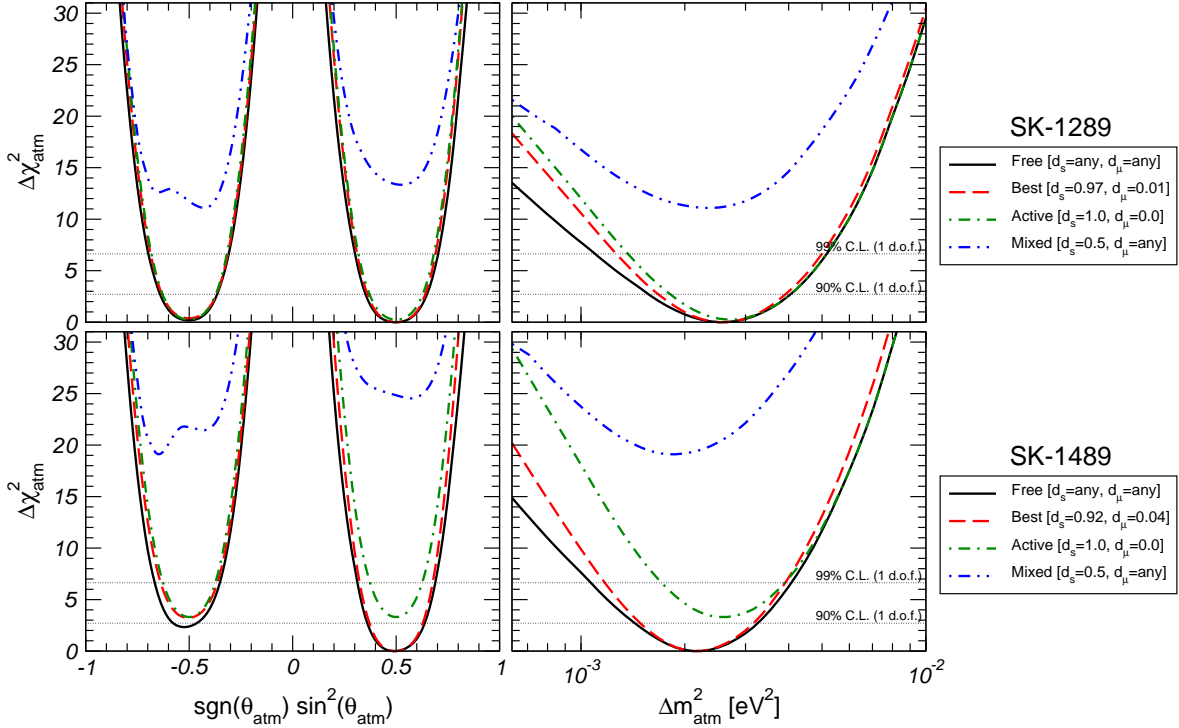}
    \caption{\label{fig:atm-chisq}%
      $\Dcq_\Atm$ as a function of $\sin^2\theta_\Atm$ (left)
      and $\Dma$ (right), using 1289 (upper) and 1489 (lower) days of
      Super-K data, for the case of neutrino oscillations with
      arbitrary $d_s$ and $d_\mu$, best-fit $d_s$ and $d_\mu$, pure
      active and mixed active-sterile neutrino oscillations.}
\end{figure}

The determination of the parameters $\theta_\Atm$ and $\Dma$ is
summarized in Figs.~\ref{fig:atm-region} and \ref{fig:atm-chisq}.  At a
given \CL\ we cut the $\chi^2_\Atm$ at a $\Dcq$ determined by 4 \dof\
to obtain 4-dimensional volumes in the parameter space of
($\theta_\Atm, \Dma, d_\mu,d_s$).  In the upper panels we show
sections of these volumes at values of $d_s=1$ and $d_\mu=0$
corresponding to the pure active case (left) and at the best-fit point
(right). Again we observe that moving from pure active to the best-fit
does not change the fit significantly.  In the lower right panel we
project away both $d_\mu$ and $d_s$, whereas in the lower left panel
we fix $d_s=0.5$ and project away only $d_\mu$.
Comparing the regions resulting from 1489 days Super-K data (shaded
regions) with those from the 1289 days Super-K sample (hollow
regions) we note that the new data leads to a slightly better
determination of $\theta_\Atm$ and $\Dma$. However, more importantly,
from the lower left panel we see, that the new data shows a stronger
rejection against a sterile admixture: for $d_s=0.5$ no allowed region
appears at 3$\sigma$ for 4 \dof.

In Fig.~\ref{fig:atm-chisq} we display the $\Dcq$ with respect to the
global best-fit point as a function of $\sin^2\theta_\Atm$ (for both
signs of $\theta_\Atm$) and $\Dma$, minimizing with respect to the
other parameter, for different assumptions on the parameters $d_s$ and
$d_\mu$. In contrast to the solar case shown in
Fig.~\ref{fig:sol-chisq} the atmospheric $\chi^2$ exhibits a beautiful
quadratic behavior, reflecting the fact that the oscillation solution
to the atmospheric neutrino problem is robust and unique.  Notice
again the significant worsening of the fit for the case of a sizable
sterile neutrino admixture (see, \eg, the line corresponding to
$d_s=0.5$).

\begin{figure}[t] \centering
    \includegraphics[width=0.95\textwidth]{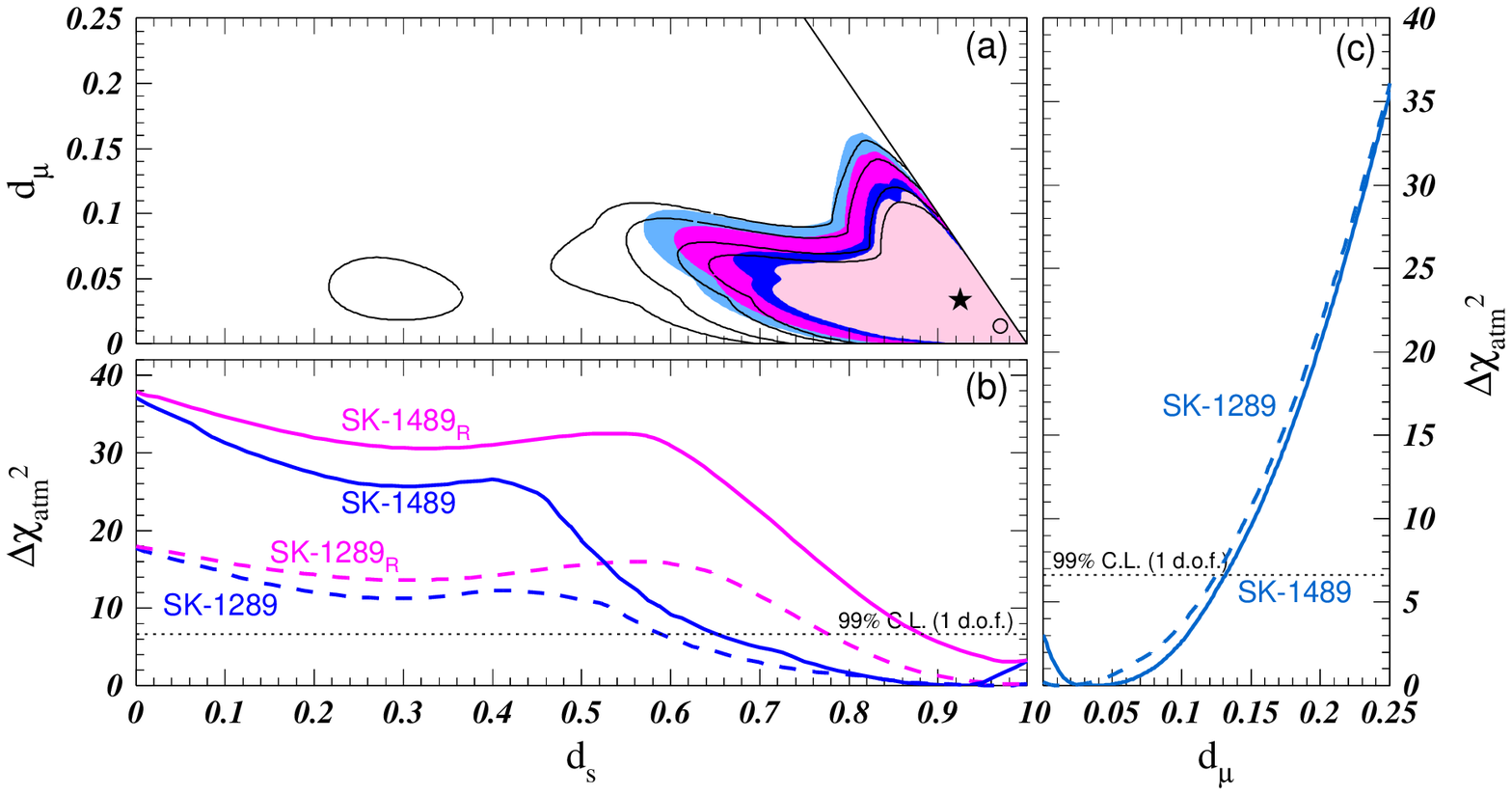}
    \caption{ \label{fig:atm-etas}%
      (a) Allowed regions of the parameters $d_s$ and $d_\mu$ at 90\%,
      95\%, 99\% and 3$\sigma$ for 2 \dof. The lines (shaded regions)
      correspond to 1289 (1489) days Super-K data. Further we show
      $\Dcq_\Atm$ as a function of $d_s$ (b) and $d_\mu$ (c),
      minimized with respect to all other parameters. The subscript
      ``R'' refers to the restricted analysis with $d_\mu=0$.}
\end{figure}

In Fig.~\ref{fig:atm-etas} we summarize the behavior of atmospheric
$\chi^2$ with respect to the parameters $d_s$ and $d_\mu$. Indeed, the
most striking result of the present improved analysis is the stronger
rejection we now obtain on the fraction of the sterile neutrino
$1-d_s$ in atmospheric oscillations.  Fig.~\ref{fig:atm-etas} (b)
clearly illustrates the degree with which the atmospheric neutrino
data sample rejects the presence of a sterile component.  On this
basis one can place a model-independent atmospheric limit on the
parameter $d_s$,
\begin{equation} \label{eq:etasAtm}
    \text{atmospheric data:} \qquad 1 - d_s \leq 0.35
\end{equation}
at 99\% \CL\ (1 \dof). For the case of the restricted analysis, in
which $d_\mu = 0$, we obtain\footnote{Note that in this case the \CL\
regions should be defined with respect to the ``restricted'' best-fit
point, which occurs for $d_s=0.99$, and not with respect to the global
one.}
\begin{equation}\label{eq:etasAtmR}
    d_\mu=0: \qquad 1 - d_s \leq 0.16 \,.
\end{equation}
By comparing Eqs.~\eqref{eq:etasAtm} and \eqref{eq:etasAtmR} we note
the importance of taking into account the finite $d_\mu$ value in the
analysis.

Although there is no substantial change in the 99\% \CL\ bounds on
$1-d_s$ due to the new Super-K data there is a huge effect for the
case of sizable sterile neutrino admixtures, $d_s \lesssim 0.5$.
In Tab.~\ref{tab:atm-chisq} we have compiled the best-fit values of
$\sin^2\theta_\Atm$, $\Dma$, the $\chi^2$ values and the GOF for the
various atmospheric data samples for pure active and pure sterile
oscillations. In the last column we give the difference in $\chi^2$
between active and sterile oscillation cases. Comparing these numbers
for the 1289 and 1489 days Super-K samples we observe that all the new
data except the Sub-GeV sample lead to a significant higher rejection
against sterile oscillations.  In combination with MACRO data the 1289
days Super-K gave a difference between pure sterile and active
oscillations of $\Dcq_\mathrm{s-a}=17.8$, whereas with the recent data
we obtain
\begin{equation}\label{eq:atm_a-s}
    \Dcq_\mathrm{s-a} = 34.6 \,,
\end{equation}
showing that pure sterile oscillations are highly disfavored with
respect to the active ones\footnote{Here we should remark that this
  big improvement in constraining the sterile component --~which is
  clearly visible also in the analyses presented by the Super-K
  collaboration itself~-- cannot be explained {\it only} by the
  improved statistics provided by the new data sample. The leading
  contribution comes instead from a change in the data themselves,
  which may indicate that some modification in the experimental
  efficiencies has been introduced. However, we have verified that
  such changes do not affect the theoretical prediction, since no
  difference between 1289 and 1489 days is visible in the Monte-Carlo
  of the Super-K collaboration.}.
Let us note that MACRO data give an important contribution to this
effect: MACRO alone disfavors the sterile oscillations already with
$\Delta \chi^2_\mathrm{s-a}=9.0$. These limits on the sterile
admixture are significantly stronger than obtained
previously~\cite{Maltoni:2001bc} and play an important role in ruling
out four-neutrino oscillation solutions in a combined global analysis
of the LSND anomaly~\cite{NewFour}.
Note, however, that in contrast with the case of $d_s$, there is no
substantial improvement in constraining the parameter $d_\mu$ due to
the new data, as seen in Fig.~\ref{fig:atm-etas} (c).

\begin{table}[t] \centering\small
    \catcode`?=\active \def?{\hphantom{0}}
    \newcommand{\D}{\cdot 10^{-3}}
    \newcommand{\Q}{\cdot 10^{-2}}
    \begin{tabular}{|>{\rule[-2mm]{0pt}{6mm}}l|c|cccc|cccc|c|}
        \hline
        & & \multicolumn{4}{c|}{Active ($d_\mu=0,\,d_s=1$)} &
        \multicolumn{4}{c|}{Sterile ($d_\mu=0,\,d_s=0$)} & \\
        \hline
        Data sample & d.o.f.
        & $\sin^2\theta$ & $\Delta m^2~[\eVq]$
        & $\chi_{\text{act}}^2$ & GOF
        & $\sin^2\theta$ & $\Delta m^2~[\eVq]$
        & $\chi_{\text{ste}}^2$ & GOF
        & $\Dcq_{\text{s--a}}$ \\
        \hline
        \multicolumn{11}{|>{\rule[-2mm]{0pt}{6mm}}c|}{Super-K-1289 days, improved} \\
        \hline
        SK Sub-GeV    & $20-2$ & $0.50$ & $2.1\D$ & $14.9$ & $67\%$ & $0.50$ & $2.2\D$ & $15.0$ & $66\%$ & $?0.1$ \\
        SK Multi-GeV  & $20-2$ & $0.50$ & $1.8\D$ & $?6.4$ & $99\%$ & $0.57$ & $3.5\D$ & $11.3$ & $88\%$ & $?4.8$ \\
        SK Stop-$\mu$ & $?5-2$ & $0.50$ & $4.2\D$ & $?1.2$ & $76\%$ & $0.61$ & $4.0\D$ & $?3.1$ & $38\%$ & $?1.9$ \\
        SK Thru-$\mu$ & $10-2$ & $0.29$ & $6.3\D$ & $?5.3$ & $73\%$ & $0.84$ & $1.0\Q$ & $?7.8$ & $45\%$ & $?2.5$ \\
        MACRO         & $10-2$ & $0.50$ & $2.4\D$ & $11.0$ & $20\%$ & $0.96$ & $9.4\D$ & $20.0$ & $?1\%$ & $?9.0$ \\
        \hline
        SK Contained  & $40-2$ & $0.50$ & $2.0\D$ & $21.4$ & $99\%$ & $0.54$ & $3.0\D$ & $26.9$ & $91\%$ & $?5.5$ \\
        Upgoing-$\mu$ & $25-2$ & $0.50$ & $3.3\D$ & $19.2$ & $69\%$ & $0.72$ & $4.2\D$ & $32.8$ & $?8\%$ & $13.6$ \\
        SK+MACRO      & $65-2$ & $0.50$ & $2.7\D$ & $41.7$ & $98\%$ & $0.56$ & $2.8\D$ & $59.4$ & $60\%$ & $17.8$ \\
        \hline
        \multicolumn{11}{|>{\rule[-2mm]{0pt}{6mm}}c|}{Super-K-1489 days} \\
        \hline
        SK Sub-GeV    & $20-2$ & $0.50$ & $1.9\D$ & $?9.0$ & $96\%$ & $0.51$ & $2.0\D$ & $?9.0$ & $96\%$ & $?0.0$ \\
        SK Multi-GeV  & $20-2$ & $0.50$ & $1.3\D$ & $10.2$ & $93\%$ & $0.57$ & $3.5\D$ & $18.4$ & $43\%$ & $?8.2$ \\
        SK Stop-$\mu$ & $?5-2$ & $0.50$ & $2.8\D$ & $?1.5$ & $69\%$ & $0.75$ & $2.8\D$ & $?6.9$ & $?8\%$ & $?5.4$ \\
        SK Thru-$\mu$ & $10-2$ & $0.50$ & $3.5\D$ & $?6.3$ & $61\%$ & $0.84$ & $6.7\D$ & $16.0$ & $?4\%$ & $?9.7$ \\
        MACRO         & $10-2$ & $0.50$ & $2.4\D$ & $11.0$ & $20\%$ & $0.96$ & $9.4\D$ & $20.0$ & $?1\%$ & $?9.0$ \\
        \hline
        SK Contained  & $40-2$ & $0.50$ & $1.5\D$ & $19.3$ & $99\%$ & $0.54$ & $3.0\D$ & $28.1$ & $88\%$ & $?8.8$ \\
        Upgoing-$\mu$ & $25-2$ & $0.50$ & $3.0\D$ & $18.9$ & $71\%$ & $0.75$ & $3.2\D$ & $40.8$ & $?1\%$ & $22.0$ \\
        SK+MACRO      & $65-2$ & $0.50$ & $2.5\D$ & $40.2$ & $99\%$ & $0.61$ & $2.7\D$ & $74.9$ & $15\%$ & $34.6$ \\
        \hline
    \end{tabular}
    \caption{ \label{tab:atm-chisq}%
      Atmospheric neutrino best-fit oscillation parameters for pure
      active and pure sterile oscillations for the various data
      samples.}
\end{table}

\begin{figure}[t] \centering
    \includegraphics[width=0.95\linewidth]{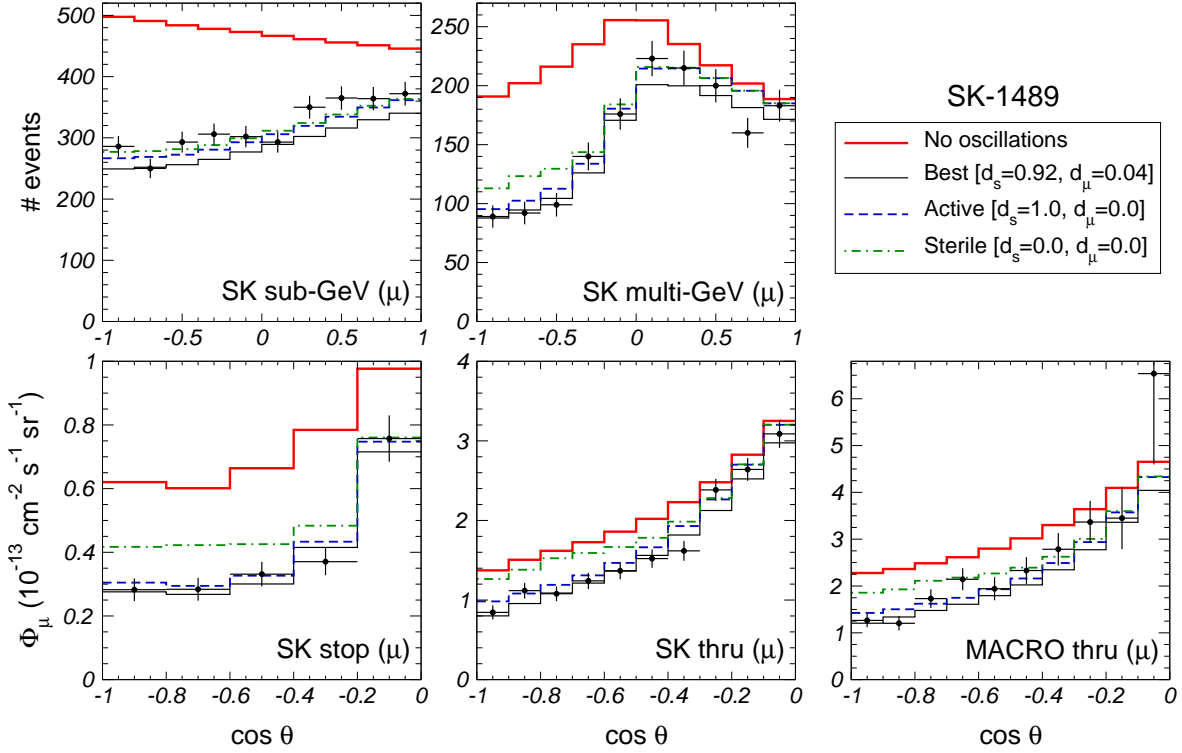}
    \caption{\label{fig:atm-zenith}%
      Zenith angle dependence of the $\mu$-like data used in our fit.
      Further we show the predicted number of atmospheric neutrino
      events for best-fit, pure-active and pure-sterile
      oscillations and no oscillations.}
\end{figure}

In order to better appreciate the excellent quality of the neutrino
oscillation description of the present atmospheric neutrino data
sample we display in Fig.~\ref{fig:atm-zenith} the zenith angle
distribution of atmospheric neutrino events.  Clearly, active neutrino
oscillations describe the data very well indeed. In contrast, no
oscillations can be visually spotted as being inconsistent. On the
other hand conversions to sterile neutrinos lead to an excess of
events for neutrinos crossing the core of the Earth, in all the data
samples except sub-GeV.


\subsection{Comparison with other groups}
\label{sec:atm-compare}

Let us briefly compare our atmospheric neutrino oscillation results
with those of other groups. Apart from the analyses presented in
Refs.~\cite{Maltoni:2001bc,concha4nu} there had been no other complete
atmospheric neutrino analysis taking into account the most general
structure of neutrino mixing in the presence of sterile neutrinos,
characterized by four mixing parameters. In the analyses of
Refs.~\cite{skatm,fogli_atm} the $\nu_\mu$ is restricted to the
atmospheric mass states, which corresponds to the constraint $d_\mu=0$
in our parameterization. However, at the corresponding limiting cases
our generalized analysis can be compared with the results of other
works.  Let us further note that the analysis of Ref.~\cite{fogli_atm}
is based on the 1289-days SK data sample (79.5 kton-yr) and in contrast to
Refs.~\cite{skatm,fogli_atm} we use also data from the MACRO
experiment.

First, we find very good agreement in the case of pure active
oscillations: the agreement of our best-fit values given in
Eq.~\eqref{eq:atmbf} with those obtained by the Super-K collaboration
($\sin^2\theta_\Atm = 0.5,\:\Dma = 2.5\times 10^{-3}~\eVq$
\cite{skatm}) is excellent, with good agreement also with the results
of Ref.~\cite{fogli_atm} ($\sin^2\theta_\Atm = 0.41,\: \Dma = 3\times
10^{-3}~\eVq$). Similarly, also the allowed ranges shown in the upper
left panel of Fig.~\ref{fig:atm-region} compare very well with the
ranges obtained in Refs.~\cite{skatm,fogli_atm}.  This shows that the
determination of the active atmospheric oscillation parameters is
already rather stable with respect to variations in the analysis and
inclusion of additional of data.
Concerning admixtures of sterile neutrinos, we note that it is
presently not possible to use information on $\nu_\tau$ appearance,
multi-ring $\mu$ and neutral-current events outside the Super-K
collaboration, because to simulate these data a detailed knowledge of
the detector and the applied cuts is necessary. These classes of
events should provide additional sensitivity towards rejecting a
possible contribution of sterile neutrinos.  Therefore, the fact that
the value of $\Dcq_\mathrm{s-a} = 49.8$~\cite{skatm} between pure
active and sterile oscillations obtained by the Super-K collaboration
is higher than our value 34.6 given in Eq.~\eqref{eq:atm_a-s} is
understandable, since with the Super-K data accessible to us we have
a reduced discrimination between active and sterile oscillations,
based solely on the matter effects.


\section{Conclusions}
\label{sec:conclusions}

Prompted by the recent data on solar and atmospheric neutrinos we have
reanalyzed the global status of oscillation solutions, taking into
account the that both the solar $\nu_e$ and the atmospheric $\nu_\mu$
may convert to a mixture of active and sterile neutrinos.
In addition to the SNO neutral current, spectral and day/night
(\snotot) results we add the latest 1496-day solar and 1489-day
atmospheric Super-K neutrino data samples.

We have studied the impact of the recent solar data in the
determination of the regions of oscillation parameters for different
allowed $\eta_s$ values, displaying the global behavior of
$\Dcq_\Sol(\Dms)$ and $\Dcq_\Sol(\theta_\Sol)$, calculated with
respect to the favored active LMA solution. We have investigated in
detail the impact of the full Cl + Ga rates + Super-K spectra + the
complete \snotot\ data set, comparing with the situation when the this
year's SNO data is left out.
We confirm the clear preference for the LMA solution of the solar
neutrino problem and obtain that the LOW, VAC, SMA and \JSQ\ solutions
are disfavored with a $\Dcq = 9$, $9$, $23$ and $31$, respectively,
for the pure active case.  In addition, we find that the global solar
data sample constrains admixtures of a sterile neutrino to be smaller
than 44\% at 99\% \CL. This bound is relaxed to 61\% when the solar
\flux[8]{B} flux is treated as a free parameter. A pure sterile
solution is ruled out with respect to the active one at 99.997\% \CL.
For allowed sterile neutrino admixtures LMA is always the best of all
the oscillation solutions. We remark, however, the existence of
non-oscillation
solutions~\cite{Miranda:2000bi,Barranco:2002te,Guzzo:2001mi}. These
will be crucially tested~\cite{Barranco:2002te,Akhmedov:2002mf} at the
up-coming KamLAND reactor experiment~\cite{kamland}.

By performing an improved fit of the atmospheric data, we have also
updated the corresponding regions of oscillation parameters for the
case where the atmospheric $\nu_\mu$ convert to a mixture of active
and sterile neutrinos. We have displayed the global behavior of
$\Dcq_\Atm(\Dma)$ and $\Dcq_\Atm(\theta_\Atm)$ for different allowed
values of the sterile neutrino admixture in the atmospheric channel.
We have compared the situation before-and-after the recent 1489-day
atmospheric Super-K data samples and shown that the GOF of the
oscillation hypothesis is excellent.
We have found that the recent 1489-day atmospheric Super-K data
strongly constrain a sterile component in atmospheric oscillations: if
the $\nu_\mu$ is restricted to the atmospheric mass states only a
sterile admixture of 16\% is allowed at 99\% \CL, while a bound of
35\% is obtained in the unconstrained case.  Pure sterile oscillations
are disfavored with a $\Dcq = 34.6$ compared to the active case.


\begin{acknowledgments}
  We thank Mark Chen, Andre Hamer, Art McDonald and Scott Oser, for
  help in accessing the details of the SNO experiment needed for our
  analysis.  This work was supported by Spanish grant BFM2002-00345,
  by the European Commission RTN network HPRN-CT-2000-00148 and by the
  European Science Foundation network grant N.~86. T.S.~has been
  supported by the DOC fellowship of the Austrian Academy of Science
  and, in the early stages of this work, by a fellowship of the
  European Commission Research Training Site contract HPMT-2000-00124
  of the host group. M.M.~was supported by contract HPMF-CT-2000-01008
  and M.A.T.\ was supported by the M.E.C.D.\ fellowship AP2000-1953.
\end{acknowledgments}


\appendix*

\section{Impact of the KamLAND result}

\begin{figure}[t] \centering
    \includegraphics[width=0.95\linewidth]{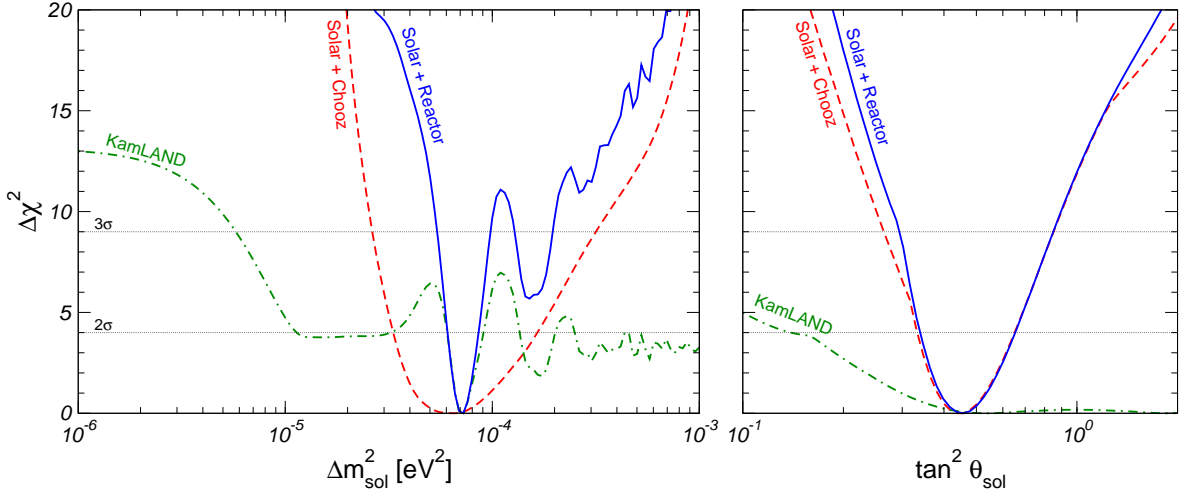}
    \caption{ \label{fig:kam-chisq}%
      $\Dcq$ as a function of $\Dms$ and $\tan^2\theta_\Sol$, for
      solar+Chooz, KamLAND alone, and solar+reactor (both Chooz and
      KamLAND) data.}
\end{figure}

In a recent paper the first results of the KamLAND collaboration
became public~\cite{kamlandPRL}.
The KamLAND experiment is a reactor neutrino experiment whose detector
is located at the Kamiokande site. Most of the $\bar{\nu}_e$ flux
incident at KamLAND comes from nuclear plants at distances of $80-350$
km from the detector, making the average baseline of about 180
kilometers, long enough to provide a sensitive probe of the LMA-MSW
region.
The target for the $\bar{\nu}_e$ flux consists of a spherical
transparent balloon filled with 1000 tons of non-doped liquid
scintillator, and the anti-neutrinos are detected via the inverse
neutron $\beta$-decay process $\bar{\nu}_e + p \to e^{+} + n$.
The KamLAND collaboration has for the first time measured the
disappearance of neutrinos produced in a power reactor.
They observe a strong evidence for the disappearance of neutrinos
during their flight over such distances, giving the first terrestrial
confirmation of the solar neutrino anomaly and also establishing the
oscillation hypothesis with man-produced neutrinos.

In this appendix (which does not appear in the published version of
this paper) we analyze the implications of the first 145.1 days of
KamLAND data on the determination of the solar neutrino parameters.
The details of our theoretical Monte-Carlo and statistical analysis
are given in Ref.~\cite{Maltoni:2002aw}; in particular, the KamLAND
$\chi^2$-function is calculated assuming a Poisson distribution for
the experimental data, as described in Sec.~IV of that paper.

\begin{figure}[t] \centering
    \includegraphics[width=0.95\linewidth]{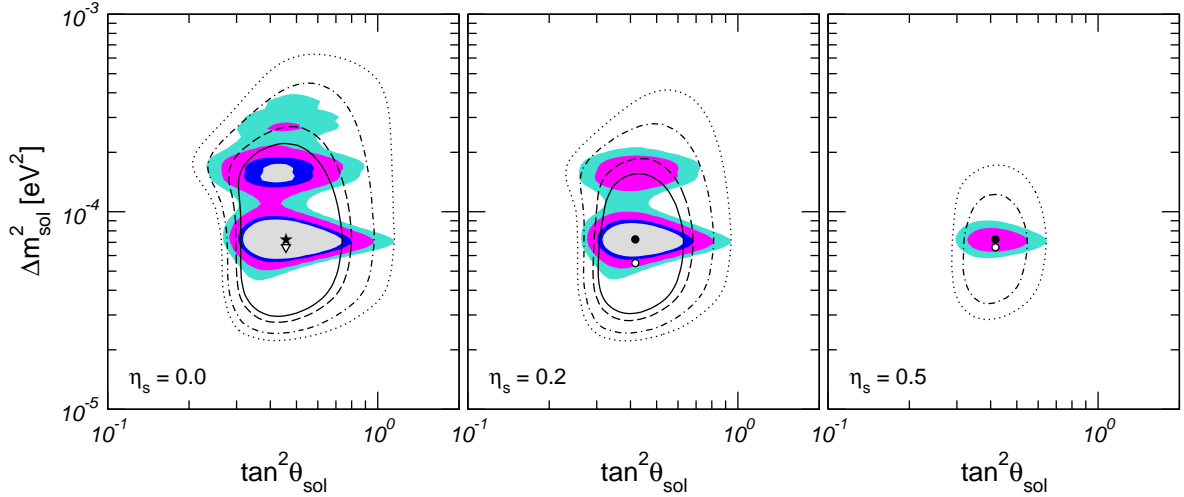}
    \caption{\label{fig:kam-region}%
      Allowed regions of $\tan^2\theta_\Sol$ and $\Dms$ for $\eta_s =
      0$ (active oscillations), $\eta_s = 0.2$ and $\eta_s = 0.5$. The
      lines correspond to the analysis of solar+Chooz data, while the
      shaded regions correspond to the combination of solar+reactor
      (both Chooz and KamLAND) data. Both lines and shaded regions
      refer to 90\%, 95\%, 99\% and 99.73\% confidence intervals for 3
      \dof.}
\end{figure}

The impact of the KamLAND result on $\theta_\Sol$ and $\Dms$ for the
case of pure active oscillations has already been discussed in detail
in Ref.~\cite{Maltoni:2002aw}, and the results are summarized here in
Fig.~\ref{fig:kam-chisq}. First of all, we note that non-LMA
solutions, characterized by a very small value of either $\Dms$ (LOW,
VAC, \JSQ) or $\theta_\Sol$ (SMA) are in disagreement with the
evidence for $\bar{\nu}_e$ disappearance observed in KamLAND. As a
consequence, the relative quality of these solutions with respect to
LMA is worsened by an extra $\Dcq \approx 13$ when the KamLAND data
are also included in the analysis, so that Eq.~\eqref{eq:dcq} is now
replaced by:
\begin{equation}
    \Dcq_{\text{LOW}}  = 21.7, \qquad
    \Dcq_{\text{VAC}}  = 21.6, \qquad
    \Dcq_{\text{SMA}}  = 36.5, \qquad
    \Dcq_{\text{\JSQ}} = 44.0.
\end{equation}
This led to the conclusion that LMA is presently the only allowed
solution to the solar neutrino problem. The global best-fit point
occurs for pure-active oscillations, and is practically unaffected by
the inclusion of KamLAND (cfr.\ Eq.~\eqref{eq:bfp-sol}):
\begin{equation} \label{eq:bfp-kam}
    \tan^2\theta_\Sol = 0.46, \qquad \Dms = 7.2\times 10^{-5}~\eVq.
\end{equation}
However, as can be seen from Fig.~\ref{fig:kam-region} the LMA region
is now split into two sub-regions, and from the left panel of
Fig.~\ref{fig:kam-chisq} we see that a secondary minimum characterized
by $\tan^2\theta_\Sol = 0.42$ and $\Dms = 1.5\times 10^{-4}~\eVq$
appears. The relative quality of this point with respect to the global
best-fit point given in Eq.~\eqref{eq:bfp-kam} is $\Dcq = 5.7$. The
allowed $3\sigma$ ranges for $\Dms$ and $\theta_\Sol$ are (cfr.\
Eq.~\eqref{eq:sol_ranges}):
\begin{equation} \label{eq:solkam_ranges}
    0.29 \le \tan^2\theta_\Sol \le 0.85, \qquad
    \left\{ \begin{aligned}
        5.4\times 10^{-5}~\eVq \le \Dms &\le 9.8\times 10^{-5}~\eVq, \\
        1.3\times 10^{-4}~\eVq \le \Dms &\le 2.0\times 10^{-4}~\eVq.
    \end{aligned}\right.
\end{equation}

Let us now consider the impact of KamLAND on the determination of
$\eta_s$. From Fig.~\ref{fig:kam-region} we have a first indication
that the bound on the fraction of sterile neutrino participating in
solar neutrino oscillations is essentially unaffected by this
experiment. This can be easily understood since KamLAND is only
sensitive to the anti-neutrino survival probability $P_{ee}$, and is
therefore unable to discriminate between different oscillation
channels. Taking into account that matter effects induced by the Earth
mantle are practically negligible given the short distance traveled
by the neutrinos in their flight between the source and the detector,
it is straightforward to conclude that KamLAND is completely
insensitive to $\eta_s$.

\begin{figure}[t] \centering
    \includegraphics[width=0.95\linewidth]{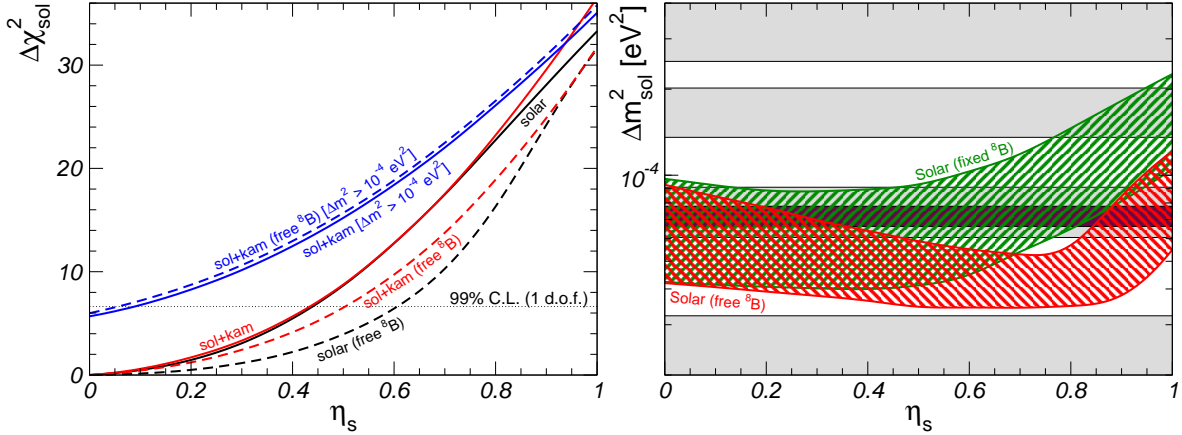}
    \caption{ \label{fig:kam-etas}%
      Left panel: $\Dcq_\Sol$ and $\Dcq_\SlKm$ displayed as functions
      of $\eta_s$, for both fixed and free \flux[8]{B} flux. Right
      panel: $1\sigma$ allowed regions for $\Dms$ from the analysis of
      solar+Chooz data, for both fixed (green) and free (red)
      \flux[8]{B} flux. The dark (light) gray horizontal belts
      correspond to the $1\sigma$ ($2\sigma)$ allowed regions for
      $\Dms$ from the analysis of KamLAND data.}
\end{figure}

In the left panel of Fig.~\ref{fig:kam-etas} we display the profile of
$\Dcq_\Sol$ and $\Dcq_\SlKm$ as functions of $\eta_s$, irrespective of
the detailed values of the solar neutrino oscillation parameters
$\Dms$ and $\theta_\Sol$. The dashed lines correspond to free
\flux[8]{B} flux, while for the solid lines the boron flux is fixed to
its SSM prediction. As expected, for the boron-fixed case the
inclusion of KamLAND is almost completely irrelevant for the
determination of the sterile neutrino fraction, and for $\eta_s
\lesssim 0.8$ no visible difference arises between the pre-KamLAND and
post-KamLAND cases. However, looking at Fig.~\ref{fig:kam-etas} we see
that for the boron-free case the inclusion of the KamLAND data is
relevant. At 99\% \CL\ we obtain the bounds (cfr.\
Eq.~\eqref{eq:etasSol}):
\begin{equation} \label{eq:etasKam}
    \text{solar+reactor data:}
    \qquad \eta_s \leq 0.43 \text{~(boron-fixed)},
    \qquad \eta_s \leq 0.51 \text{~(boron-free)}.
\end{equation}

To understand why fixing or not-fixing the \flux[8]{B} flux leads to
such a different behavior, we illustrate in the right panel of
Fig.~\ref{fig:kam-etas} the dependence on $\eta_s$ of the $1\sigma$
allowed range for $\Dms$ from the analysis of solar data alone, both
for the boron-fixed (green) and the boron-free (red) cases. These two
bands should be confronted with the horizontal gray
belts\footnote{Note that the gray bands are perfectly horizontal since
  KamLAND alone is insensitive to $\eta_s$.}, which correspond to the
KamLAND $1\sigma$ (dark gray) and $2\sigma$ (light gray) allowed
intervals for $\Dms$.

\begin{figure}[t] \centering
    \includegraphics[width=0.45\linewidth]{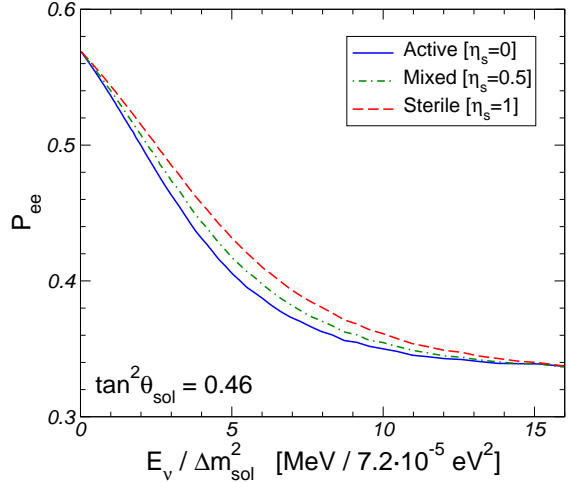}
    \caption{ \label{fig:kam-probab}%
      Solar neutrino survival probability $P_{ee}$ as a function of
      $E_\nu/\Dms$, calculated at the solar+reactor best-fit point
      given in Eq.~\eqref{eq:bfp-kam}.}
\end{figure}

Concerning the boron-fixed case, we see that for $\eta_s \lesssim 0.6$
the $\Dms$ value preferred by solar data is almost insensitive to
$\eta_s$, and in very good agreement with the KamLAND prediction. In
this regime a non-zero value of $\eta_s$ leads to a mild deficit in
the expected number of NC and ES events, thus reducing the quality of
the fit. When $\eta_s$ exceeds $\sim 0.6$, this deficit become
relevant, and in order to compensate it the best-fit point moves
towards regions of the neutrino parameter space where the electron
neutrino survival probability $P_{ee}$ is larger. From
Fig.~\ref{fig:kam-probab} it is easy to understand that this
correspond to larger values of $\Dms$. In any case, as long as $\eta_s
\lesssim 0.9$ the $\Dms$ regions allowed at $1\sigma$ by solar and
KamLAND data still overlap, and this explains why in this regime there
is no visible difference between $\Dcq_\Sol$ and $\Dcq_\SlKm$.

The situation is different for the boron-free case. As for the
previous case, an increase of $\eta_s$ lead to a deficit in the
expected number of NC and ES events, which can now be compensated by
assuming a larger value of the \flux[8]{B} flux. However, a larger
boron flux results in an excess of $\nu_e$ arriving at the detectors,
so that now the experimental data favor regions of the parameter space
where the electron neutrino survival probability $P_{ee}$ is smaller.
This explains why the preferred value for $\Dms$ decreases as $\eta_s$
increases. When $\eta_s \gtrsim 0.8$ the favored value for the
\flux[8]{B} flux rapidly decreases, and $\Dms$ increases again.
A consequence of this is that already for $\eta_s \gtrsim 0.4$ the
$1\sigma$ regions for solar and for KamLAND data no longer overlap.
This leads to a tension between the two data sets, which results in an
excess of $\Dcq_\SlKm$ over $\Dcq_\Sol$. It is only for very large
values of $\eta_s$ that solar data return in agreement with KamLAND,
and the two lines $\Dcq_\Sol$ and $\Dcq_\SlKm$ merges again.

In summary, in this appendix we have investigated the impact of the
first 145.1 days of KamLAND data on the determination of the solar
neutrino parameters. We have found that all non-LMA solution are now
ruled out, and that the original LMA region is split into two
relatively narrow islands around the values of $\Dms = 7.2\times
10^{-5}~\eVq$ (best fit point) and $\Dms = 1.5\times 10^{-4}~\eVq$
(local minimum). The bound on the sterile neutrino fraction $\eta_s$
is practically unaffected in the boron-fixed case, but improves from
$0.61$ to $0.51$ (at 99\% \CL) in the boron-free case.

\end{document}